\documentclass[aps,twocolumn,pra,superscriptaddress,amsmath,amssymb]{revtex4-1}
\usepackage{graphicx,color}
\usepackage{epstopdf}
\usepackage{epsfig}
\usepackage[breaklinks=true]{hyperref}

\newcommand{\beq}{\begin{equation}}
\newcommand{\eeq}{\end{equation}}
\newcommand{\bqa}{\begin{eqnarray}}
\newcommand{\eqa}{\end{eqnarray}}

\newcommand{\erf}[1]{Eq.~(\ref{#1})}

\newcommand{\smallfrac}[2]{\mbox{$\frac{#1}{#2}$}}
\newcommand{\bra}[1]{\left\langle{#1}\right|}
\newcommand{\ket}[1]{\left|{#1}\right\rangle}

\newcommand{\op}[2]{\ket{#1}\!\bra{#2}}
\newcommand{\sch}{Schr\"odinger}

\newcommand{\half}{\smallfrac{1}{2}}

\newcommand{\sq}[1]{\left[ {#1} \right]}

\newcommand{\tr}[1]{{\rm Tr}\sq{ {#1} }}

\definecolor{green}{rgb}{0.00,0.50,0.00}

\begin{document}
\title{Quantum Filtering (Quantum Trajectories) for Systems Driven by Fields in 
Single Photon and Superposition of Coherent States}

\author{John E.~Gough} \email{jug@aber.ac.uk}
   \affiliation{Institute for Mathematics and Physics, Aberystwyth University, SY23 3BZ, Wales, United Kingdom}
\author{Matthew R.~James}\email{Matthew.James@anu.edu.au}
  \affiliation{ARC Centre for Quantum Computation and Communication Technology}
  \affiliation{Research School of Engineering, Australian National University, Canberra, ACT 0200, Australia}
 \author{Hendra I.~Nurdin} \email{School of Electrical Engineering and Telecommunications, The University of New South Wales, Sydney, NSW 2052, Australia; h.nurdin@unsw.edu.au}
  \affiliation{Research School of Engineering, Australian National University, Canberra, ACT 0200, Australia}
\author{Joshua Combes}\email{combes@unm.edu}
  \affiliation{Research School of Engineering, Australian National University, Canberra, ACT 0200, Australia}
   \affiliation{Department of Physics and Astronomy, University of New Mexico, Albuquerque NM 87131-0001, USA}

\date{\today}

\begin{abstract}
We derive the stochastic master equations, that is to say, quantum filters, and master equations for an arbitrary quantum system probed by a continuous-mode bosonic input field in two types of non-classical states. Specifically, we consider the cases where the state of the input field is a superposition or combination of: (1) a continuous-mode single photon wave packet and vacuum, and (2) any continuous-mode coherent states. 
\end{abstract}

\maketitle

\section{Background and Motivation}
\label{sec:Intro}
The production and verification of non-classical states of light, such as
single-photon states \cite{LvoAicBen01} and superpositions of coherent
states (also known as Schr\"odinger\ cat states) \cite
{NeeNieHet06,OurTuaLau06,OurHyuTua07}, has become routine. In particular,
the production of single photon states has been achieved in a variety of
experimental architectures such as: cavity quantum electrodynamics (QED) 
\cite{KuhHenRem02,McKBocBoo04}, quantum dots in semiconductors \cite
{YuaKarSte02}, and recently in circuit QED~\cite{jay}. Such non-classical
states have been considered in connection with quantum computing \cite
{KLM01,RalGilMil03} and secure communication \cite{GisRibTit02} over quantum
networks \cite{CirZolKim97}.

\begin{figure}[h]
\begin{center}
\includegraphics[scale=0.7]{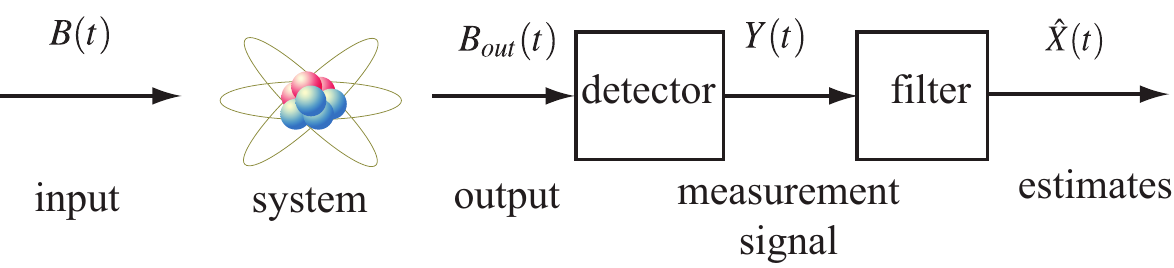} 
\end{center}
\caption{A schematic representation of a continuous measurement process, where the  measurement  signal   produced by a detector is filtered to produce estimates $\hat X(t)=\protect\pi_t(X)=\mathrm{tr}[ \protect\rho(t)
X] $ of system operators $X$ at time $t$.}
\label{fig:system-filter}
\end{figure}

A basic problem in quantum optics concerns the extraction of information about a system of interest (two-level atom, cavity mode, etc)  from light scattered by the system, Figure \ref{fig:system-filter}. Based on measurements of the scattered, or output, light, one can determine a conditional state from which one can make estimates of observables of the system. A general approach to estimation problems of this kind, called {\em filtering} problems, was developed by Belavkin \cite{Belavkin1}-\cite{BarBel91} within a framework of continuous non-demolition quantum measurement
in the case where the input probe field, $B(t)$ in Figure \ref{fig:system-filter}, is a quantum white noise with
vacuum state (or more generally Gaussian state, see \cite{GSobolev04}- \cite{GK10}).  Belavkin's formulation, which generalizes the classical nonlinear filtering theory \cite{Stratonovich}, 
is quite general. For example, 
in the schematic representation of a continuous measurement process shown in Figure \ref{fig:system-filter}, the measurement  signal  $Y(t)$ produced by a detector (e.g. photon counter or homodyne detector)  may be  the number of quanta in the output field, or alternatively it
may be a quadrature of the output field.

One obtains a filtering
equation which is a stochastic differential equation of the state $\varrho (t)$
conditioned on $Y(t)$: in the later terminology employed in quantum optics,
the output is referred to as a quantum trajectory and the filtering equation
as a stochastic master equation \cite{CarBook93,DalCasMol92,GisPer92}.
Averaging over the measured output is equivalent to a non-selective
measurement, and the corresponding state will satisfy the corresponding master
equation. The choice of detection scheme on the output field determines the
particular selective evolution, usually referred to as an unravelling of the master equation in quantum optics. To date,
quantum trajectories and quantum filtering have only been developed for
input fields that are in a Gaussian state, with specific cases being
coherent state fields (this includes vacuum fields as special case), thermal
fields, and squeezed fields \cite{GarCol85,DumParZol92,GarZol00}.
While the resulting equations allow us to estimate non-commutating observables of the
monitored system, the Gaussian nature of the inputs ensure that they appear
formally similar to the classical equations. The aim of the present paper is
to extend the theory to classes of non-classical inputs.

In this article we extend Belavkin's quantum filtering theory and the input-output theory of quantum optics \cite{GarCol85,CarBook93} to
non-Gaussian continuous-mode states $\rho_{\text{field}}$ which are superpositions or combinations of

\begin{itemize}
\item[a) ]  a continuous-mode single photon and vacuum, and

\item[b) ]  continuous-mode coherent states, i.e. continuous-mode cat-states.
\end{itemize}

The problem to be tackled here is to derive the master and stochastic master
equations for a ``system''$\otimes $``field'' with initial state $\rho
_{0}\otimes \rho _{\text{field}}$ and unitary evolution process $U\left(
t\right) $ when the field state is one of the  states above. To make the problem tractable, we seek a
larger representation of the form ``extended system''$\otimes $``field'' where 
\begin{equation*}
\text{``extended system''=``ancilla''}\otimes \text{``system''}
\end{equation*}
such that, for all system observables $X$, 
\begin{align}
\text{\textrm{tr}}_{\text{system}\otimes \text{field}} \left\{ \rho _{0}\otimes
\rho _{\text{field}}\,U\left( t\right) ^{\dag }\left( X\otimes I\right)
U\left( t\right) \right\} & \notag \\
 = \text{\textrm{tr}}_{\text{ancilla}\otimes \text{system}%
\otimes \text{field}} \big\{ \rho_a \otimes \rho _{0}\otimes P_{%
\text{vac}} & \notag \\
\tilde{U}\left( t\right) ^{\dag } \left(  R\left( t\right)
\otimes X\otimes I\right) \tilde{U}\left( t\right) \big\} &
\label{eq:basic} 
\end{align}
where $\tilde{U}\left( t\right) $ is a unitary evolution process coupling the ancilla, system, and field, $\rho_a$ 
is a fixed state of the ancilla, $P_{\text{vac}}=|0 \rangle \langle 0|$ is the vacuum state (projection) for the continuous-mode field, and $R\left( t\right) $ is some
process taking values in the observables of the ancilla. The
filtering problem may then be solved for the extended system with reference
to the vacuum state for the field using traditional techniques.

The extension to single photon states is interesting for foundational
reasons \cite{TanWalCol91} as well as the aforementioned technological
reasons \cite{KLM01}. Likewise, quantum filtering for cat states is of
foundational importance, while practical uses would be towards quantum
enhanced metrology \cite{WisMilBook}. One possible application would be to
quantum enhanced metrology of a time varying parameter \cite
{MilMunNem02,MunRalGla04}.

This article is structured as follows. In Section~\ref{sec:open} we review
standard input-output theory. Specifically we consider the idealized quantum
white-noise model of a quantum stochastic differential equation (QSDE) and
use it to derive the master equations and quantum trajectories for Gaussian
fields. Then we review a general parametrization to specify the system
environment coupling for input-output systems. Using this parametrization we
review the methods, recently introduced \cite{YanKim03_1,GouJam09a,GouJam09b}%
, to simplify and formalize the network theory of cascaded open quantum
systems {and quantum feedback networks}.

Section~\ref{sec:photon} is focused on deriving the master equation and
stochastic master equation (quantum filter) driven by continuous-mode single
photon wave packets. We generalize the single photon filter to any
superposition or combination of single photon and vacuum input field. 
The system that is probed is left arbitrary so in general our filter can
apply to qubits, qudits and mechanical oscillators. As an example we
calculate the single photon filter for a two level atom (or qubit)
dispersively coupled to the field. We derive the trajectories for both a
homodyne type measurement and a photon counting measurement.

In Section~\ref{sec:cat} we present the extension to superpositions of
coherent states. We derive the cat-state-filter for an arbitrary quantum system and an
arbitrary cat state. Again we illustrate the filtering equations with a
qubit system and homodyne and photon counting measurements.

In Section~\ref{sec_conc} we conclude and discuss our future research and
some open questions.

\noindent \textbf{Notation} 
The commutator and anti-commutator will be denoted as $[A,B]=AB-BA$ and $%
[A,B]_{+}=AB+BA$, respectively. We set $\mathcal{D}_{A}B\equiv A^{\dag }BA-%
\frac{1}{2}(A^{\dag }AB+BA^{\dag }A)$ and $\mathcal{D}_{A}^{\star }B\equiv
ABA^{\dag }-\frac{1}{2}(A^{\dag }AB+BA^{\dag }A)$.

The scattering, coupling and Hamiltonian operators describing a given
Markovian open system coupling will be written as a triple $G=(S,L,H)$, to be explained
in more detail in Section \ref{sec:open-io}, and
this provides an operator-valued parameterization of the system. The
associated superoperators are 
\begin{align*}
\mathit{Lindbladian:}& \,\mathcal{L}_{G}X\equiv -i[X,H]+\mathcal{D}_{L}X, \\
\mathit{Liouvillian:}& \,\mathcal{L}_{G}^{\star }\rho \equiv -i[H,\rho ]+%
\mathcal{D}_{L}^{\star }\rho ,
\end{align*}
and note that, for traceclass $\rho $ and bounded $X$, 
\begin{equation*}
\mathrm{tr}\{\rho \mathcal{L}_{G}X\}=\mathrm{tr}\{X\mathcal{L}_{G}^{\star
}\rho \}.
\end{equation*}


\section{Models of Open Quantum Systems}

\label{sec:open}
In this section we briefly review quantum stochastic calculus (input-output
theory) and quantum filtering (trajectories) for a system coupled to a heat
bath modelled as a boson field in the vacuum state.


\subsection{Input-Output Model Using QSDEs}

\label{sec:open-io} 

Hudson and Parthasarathy \cite{HudPar84,Par92} showed
how to dilate a dissipated completely positive semigroup evolution, with
Lindblad generator, to a unitary model on the system space with a (Bose)
Fock space ancilla. Here they developed an analogue to the It\={o} theory of
stochastic integration with respect to creation, annihilation and scattering
process $B^{\dag }\left( t\right) ,B\left( t\right) $ and $\Lambda \left(
t\right) $. They showed the existence and uniqueness of solutions to unitary
quantum stochastic differential equations of the form

\begin{align}
dU(t)=&\biggl\{ \left( S-1\right) d\Lambda \left( t\right) +LdB^{\dag }(t)  \notag
\\
&-L^{\dag }SdB(t)-(\half L^{\dag }L+iH)dt\biggr\}U(t).
\label{eq:open-schrodinger-1}
\end{align}
where 
\[ 
G=\left( S,L,H\right)
\]
consists of a unitary $S$ describing photon scattering phase, a bounded operator 
$L$ describing coupling to the creation mode of the field, and a bounded Hermitean operator $H$ describing the system Hamiltonian. (The
result has been extended to non-bounded coefficients.) The increments are
future pointing operator-valued It\={o} increment, that is $dB(t)\equiv
B(t+dt)-B(t)$ is a forward of the quantum noise. In particular, we have. $%
[U(t),dB(t)]=[U(t),dB^{\dag }(t)]=[U(t),d\Lambda(t)]=0$. The full quantum It\={o} table is 
\begin{equation}
\begin{tabular}{l|llll}
$\times $ & $dt$ & $dB$ & $d\Lambda $ & $dB^{\dag }$ \\ \hline
$dt$ & 0 & 0 & 0 & 0 \\ 
$dB$ & 0 & 0 & $dB$ & $dt$ \\ 
$d\Lambda $ & 0 & 0 & $d\Lambda $ & $dB^{\dag }$ \\ 
$dB^{\dag }$ & 0 & 0 & 0 & 0
\end{tabular} .
\label{eq:vac_ito}
\end{equation}
More generally, for quantum stochastic integral processes $X(t),Y(t)$, one
has the It\={o} product rule 
\begin{equation*}
d (X (t) Y(t) ) =( dX(t)) \, Y(t) + X(t) \, dY(t) +dX(t) \, dY(t).
\end{equation*}

Independently, Gardiner and Collett developed an equivalent quantum
input-output theory \cite{GarCol85,GarZol00} based on
Lehmann-Symanzik-Zimmermann scattering theory of Bose white noise processes.
Formally one begins with singular fields satisfying 
\begin{equation*}
\left[ b\left( t\right) ,b^{\dag }\left( s\right) \right] =\delta \left(
t-s\right) ,
\end{equation*}
with the connection to the regular processes being formally 
\begin{eqnarray*}
B^{\dag }\left( t\right) &=&\int_{0}^{t}b^{\dag }\left( s\right) ds,\quad
B\left( t\right) =\int_{0}^{t}b\left( s\right) ds, \\
\Lambda \left( t\right) &=&\int_{0}^{t}b^{\dag }\left( s\right) b\left(
s\right) ds.
\end{eqnarray*}
The quantum stochastic calculus may then be understood as effectively
arising through Wick ordering of the singular fields.

The multiple input version is relatively straightforward. We have $n$ independent
inputs $b_{j}$ and with $B_{j}\left( t\right) =\int_{0}^{t}b_{j}\left(
s\right) ds$, $\Lambda _{jk}\left( t\right) =\int_{0}^{t}b_{j}^{\dag }\left(
s\right) b_{k}\left( s\right) ds$, etc., we have 
\begin{align*}
dU(t) &=\left\{ \sum_{jk}\left( S_{jk}-\delta _{jk}\right) d\Lambda
_{jk}\left( t\right) +\sum_{j}L_{j}dB_{j}^{\dag }(t)\right. \\
& \left. -\sum_{jk}L_{j}^{\dag }S_{jk}dB_{k}(t)-\biggl(\frac{1}{2}%
\sum_{j}L_{j}^{\dag }L_{j}+iH\biggr)dt\right\} U(t),
\end{align*}
where we now have parameterizing operators 
\begin{equation*}
S=\left( 
\begin{array}{ccc}
S_{11} & \dots & S_{1n} \\ 
\vdots & \ddots & \vdots \\ 
S_{n1} & \dots & S_{nn}
\end{array}
\right) ,L=\left( 
\begin{array}{c}
L_{1} \\ 
\vdots \\ 
L_{n}
\end{array}
\right) ,H
\end{equation*}
with $S$ unitary and $H$ self-adjoint. For simplicity we treat the case of a
single input and output.

\subsection{Heisenberg-Langevin Equations}

The Heisenberg dynamics of arbitrary system operator $X$ is defined by
transforming to the Heisenberg picture 
\[
j_t (X) = U^\dag(t) (X \otimes I_{\mathrm{field}}) U(t).
\]
(We  will usually drop the subscripts ``system''
and ``field' when there is no confusion.) From the quantum It\={o} product rule and
table one deduces the QSDE for a system operator $j_t (X) = X(t)$: with all
system operators transformed to the Heisenberg picture. 
\begin{align}
dj_t(X) &= j_t ( \mathcal{L} X)dt  \notag \\
& + dB^\dag (t)j_t (S^\dag[X,L] ) + j_t ([L^\dag, X] S)dB(t)  \notag \\
& + j_t(S^\dag X S - X) d\Lambda(t).  \label{eq:qsde-X-1-S}
\end{align}


\subsection{Derivation of the Master Equation}

\label{sec:open-master} 

Suppose that the system is in an initial state $\rho (0)=\rho _{0}$ and that
the joint state of the system and bath is $\rho _{0}\otimes P_{\text{vac}}$
where $P_{\text{vac}}=|0\rangle \langle 0|$ is projection onto the vacuum
state of the field. The state of the system, $\varrho(t)$, obtained by averaging over the
environment at a given time $t$ is then 
\begin{equation}
\varrho (t)=\mathrm{tr}_{\text{field}}\left[ U(t)(\rho _{0}\otimes P_{\text{vac}%
})U^{\dag }(t)\right] .  \label{eq:master-sigma-2}
\end{equation}
We wish to obtain a differential equation for the average of an observable $%
X $ of the system at time $t$: 
\begin{eqnarray*}
\varpi _{t}\left( X\right) &=&\mathrm{tr}_{\text{system}\otimes \text{field}%
}\left\{ j_{t}\left( X\right) \varrho _{0}\otimes P_{\text{vac}}\right\} \\
&\equiv &\mathrm{tr}_{\text{system}}\left\{ \rho \left( t\right) X\right\} ,
\end{eqnarray*}
and from the Heisenberg-Langevin equation (\ref{eq:qsde-X-1-S}) we have 
\begin{eqnarray*}
d\varpi _{t}\left( X\right) &=&\mathrm{tr}_{\text{system}\otimes \text{field}%
}\left\{ dj_{t}\left( X\right) \rho _{0}\otimes P_{\text{vac}}\right\} \\
&=&\mathrm{tr}_{\text{system}\otimes \text{field}}\left\{ j_{t}\left( 
\mathcal{L}_{G}X\right) \rho _{0}\otimes P_{\text{vac}}\right\} dt,
\end{eqnarray*}
as the increments $dB,dB^{\dag },d\Lambda $ vanish in the vacuum state. We
therefore obtain the equation 
\begin{equation*}
\frac{d\varpi _{t}\left( X\right) }{dt}=\varpi _{t}\left( \mathcal{L}%
_{G}X\right) ,\quad \varpi _{0}\left( X\right) =\mathrm{tr}_{\text{system}%
}\left\{ \rho _{0}X\right\}
\end{equation*}
which may then be expressed as the master equation 
\begin{equation}
\frac{d\varrho (t)}{dt}=\mathcal{L}_{G}^{\star }\varrho(t) \equiv -i[H,\varrho (t)]+%
\mathcal{D}_{L}^{\star }\varrho (t),  \label{eq:master-1}
\end{equation}
with initial data $\rho _{0}$. Note that the master equation (\ref
{eq:master-1}) is a consequence of the QSDE model.


\subsection{The Input-Output Relations}

\label{sec:ior} 

The output field $B_{\text{out}}$ is obtained from the input by moving into
the Heisenberg picture: 
\begin{eqnarray*}
B_{\text{out}}\left( t\right) &=&U\left( t\right) ^{\dag }\left( I_{\text{%
system}}\otimes B\left( t\right) \right) U\left( t\right) \\
&\equiv &U\left( \tau\right) ^{\dag }\left( I_{\text{system}}\otimes B\left(
t\right) \right) U\left( \tau \right)
\end{eqnarray*}
for any $\tau \geq t$. Again from the quantum It\={o} calculus we find 
\begin{equation}
dB_{\text{out}}\left( t\right) =j_{t}\left( S\right) dB\left( t\right)
+j_{t}\left( L\right) dt.
\end{equation}
Note that the output field again satisfies the canonical commutation
relations.

\begin{figure}[h!]
\begin{center}
\includegraphics[scale=0.7]{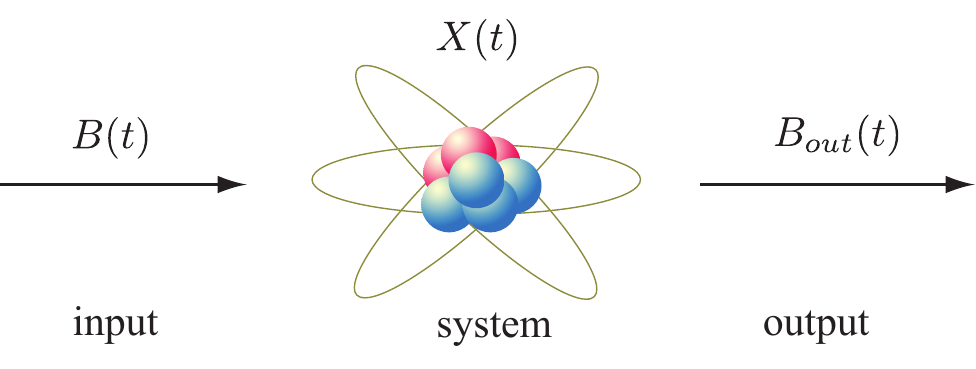} 
\end{center}
\caption{An open quantum system. The input field (before interaction) is
represented by the operator $B(t)$ and output field (after interaction) is
denoted by $B_{\mathrm{out}}(t)$. }
\label{fig:open1}
\end{figure}


\subsection{Derivation of the Quantum Filter (Stochastic Master Equation) -
Quadrature Case}

\label{sec:open-trajectory} 

We suppose that we continuously monitor the quadrature phase using perfect 
(100\% efficiency) homodyne detection. This entails measurement, for each $t\geq 0$%
,  of the field 
\begin{eqnarray*}
Y(t) &=&B_{\mathrm{out}}(t)+B_{\mathrm{out}}^{\dag }(t) \\
&\equiv &U\left( t\right) ^{\dag }\left( I_{\text{system}}\otimes Q\left(
t\right) \right) U\left( t\right)
\end{eqnarray*}
where $Q\left( t\right) =B\left( t\right) +B^{\dag }\left( t\right) $. We
note that the set of observables $\left\{ Y\left( t\right) :t\geq 0\right\} $
is self-commuting and we may simultaneously diagonalize (and measure!) all
observables. At any time $t$, we may additionally estimate an observable
that commutes with the observables up to time $t$. This includes observables 
$X\left( \tau \right) $ for $\tau \geq t$, since 
\begin{eqnarray*}
&&\left[ X\left( \tau \right) ,Y\left( t\right) \right] \\
&=&U\left( \tau \right) ^{\ast }\left[ X\otimes I_{\text{field}},I_{\text{system}%
}\otimes Q\left( t\right) \right] U\left( \tau \right) \\
&\equiv &0\text{.}
\end{eqnarray*}
This is the non-demolition property. Quantum filtering is the estimation of $%
j_{t}\left( X\right) $ based on observations of the output processes$\left\{
Y\left( s\right) :0\leq s\leq t\right\} $. Fig.~\ref{fig:system-filter}
depicts the scenario we are considering. From the It\={o} calculus we see
that 
\begin{equation*}
dY(t)=(L(t)+L^{\dag }(t))dt+dQ(t).
\end{equation*}

Defining the expectation 
\begin{equation*}
\mathbb{E}[\cdot ]=\mathrm{tr}\{\rho _{0}\otimes \rho _{\text{field}%
}\,(\cdot )\}
\end{equation*}
for a given state $\rho _{0}\otimes \rho _{\text{field}}$, we seek to
minimize 
\begin{equation*}
\mathbb{E}[(\hat X(t)-j_{t}\left( X\right) )^{2}]
\end{equation*}
over all observables $\hat{X}\left( t\right) $ in the algebra $\mathcal{Y}%
_{t}$ generated by $\left\{ Y\left( s\right) :0\leq s\leq t\right\} $. The
minimizer is called the least-squares estimator for $X\left( t\right) $
given $\left\{ Y\left( s\right) :0\leq s\leq t\right\} $ and will be denoted
as 
\begin{equation}
\hat{X}(t)=\pi _{t}(X)=\mathbb{E}[j_{t}(X)\,|\,\mathcal{Y}_{t}].
\end{equation}
The later notation suggest that in $\pi _{t}\left( X\right) $ is the
conditional expectation of $j_{t}\left( X\right) $ given the past history,
which would be the classical interpretation. While conditional expectations
generally do not exist in the quantum probabilistic setting, the
nondemolition property above suffices to allow one to realize precisely this
interpretation, see for instance \cite{Bel94, BouvanHJam07}. The conditional
expectation can indeed be interpreted as an orthogonal projection onto a
subspace of commuting operators $\mathcal{Y}_{t}$. This means that $%
j_{t}(X)-\pi _{t}(X)$ is orthogonal to this measurement subspace $\mathcal{Y}%
_{t}$, that is, 
\begin{equation}
\mathbb{E}[(j_{t}(X)-\pi _{t}(X))C]=0  \label{ortho}
\end{equation}
for all operators $C$ belonging to the measurement subspace $\mathcal{Y}_{t}$%
, \cite{BouvanHJam07}. Setting $C=I$ shows that 
\begin{equation*}
\mathbb{E}[\pi _{t}(X))]=\mathbb{E}[j_{t}(X)].
\end{equation*}
\begin{widetext}
\begin{center}

\begin{figure}[h]
\includegraphics[width= 0.9\linewidth]{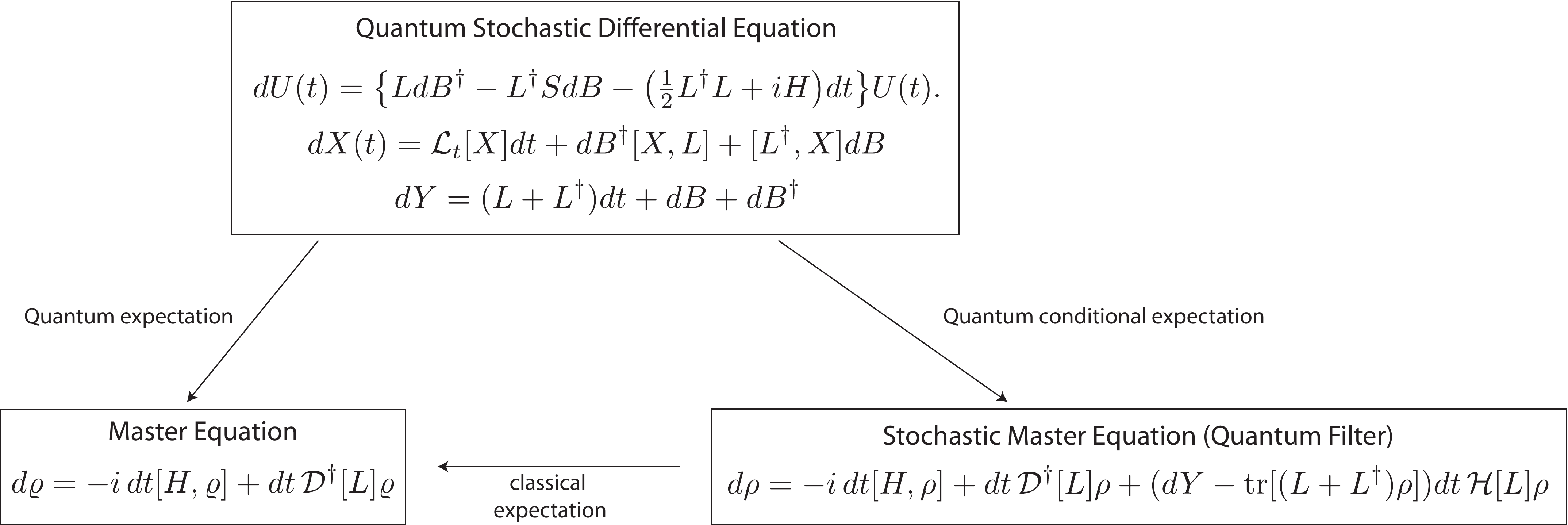}
\caption{\label{fig:unravel} Relationship between QSDEs for the unitary, system and environment; master equation; and stochastic master equation (filter) for open quantum systems (${\mathcal H}_L$ is given by (\ref{eq:calH}). The difference between the master equation and stochastic master equation is due to the difference in the type of expectation take, i.e. unconditioned or conditioned, respectively.}
\end{figure}
\end{center}
\end{widetext}

Now let us return to the vacuum state for the field: $\rho _{\text{field}%
}=P_{\text{vac}}$. We shall recall a simple derivation of the filter using
an analogue of a the characteristic function technique of classical
filtering \cite{AM}. We introducing a process $C\left( t\right) $ satisfying
the QSDE 
\begin{equation}
dC\left( t\right) =g\left( t\right) C\left( t\right) dY\left( t\right) ,
\end{equation}
with initial condition $C\left( 0\right) =I$. Here we assume that $g$ is
integrable, but otherwise arbitrary. The technique is to make an ansatz of
the form 
\begin{equation}
d\pi _{t}\left( X\right) =\alpha _{t}dt+\beta _{t}\left( X\right) dY\left(
t\right)  \label{filter ansatz}
\end{equation}
where we assume that the processes $\alpha _{t}$ and $\beta _{t}$ are
adapted and lie in $\mathcal{Y}_{t}$. These coefficients may be deduced from
the identity 
\begin{equation*}
\mathbb{E}\left[ \left( \pi _{t}\left( X\right) -j_{t}\left( X\right)
\right) C\left( t\right) \right] =0
\end{equation*}
which is valid since $C(t)$ is in $\mathcal{Y}_{t}$. We note that the
It\={o} product rule implies $I+II+III=0$ where

\begin{eqnarray*}
I &=&\mathbb{E}\left[ \left( d\pi _{t}\left( X\right) -dj_{t}\left( X\right)
\right) C\left( t\right) \right] , \\
&=&\mathbb{E}\left[ \alpha _{t}C\left( t\right) +\beta _{t}j_{t}\left(
L+L^{\dag }\right) C\left( t\right) \right] dt \\
&&-\mathbb{E}\left[ j_{t}\left( \mathcal{L}_{G}X\right) C\left( t\right) %
\right] dt, \\
II &=&\mathbb{E}\left[ \left( \pi _{t}\left( X\right) -j_{t}\left( X\right)
\right) dC\left( t\right) \right] , \\
&=&\mathbb{E}\left[ \left( \pi _{t}\left( X\right) -j_{t}(X)\right) g\left(
t\right) C\left( t\right) j_{t}\left( L+L^{\dag }\right) \right] dt, \\
III &=&\mathbb{E}\left[ \left( d\pi _{t}\left( X\right) -dj_{t}\left(
X\right) \right) dC\left( t\right) \right] \\
&=&\mathbb{E}\left[ \beta _{t}g\left( t\right) C\left( t\right) \right] dt+ \mathbb{E} \left[ 
g(t) j_t ( [L^\dag ,X ])  C(t) \right] dt.
\end{eqnarray*}

Now from the identity $I+II+III=0$ we may extract separately the
coefficients of $g\left( t\right) C\left( t\right) $ and $C\left( t\right) $
as $g\left( t\right) $ was arbitrary to deduce 
\begin{eqnarray*}
\pi _{t}\left( \left( \pi _{t}\left( X\right) -j_{t}(X)\right) j_{t}\left(
L+L^{\dag }\right) \right) +\pi _{t}\left( \beta _{t}\right) &=&0, \\
\pi _{t}\left( \alpha _{t}+\beta _{t}j_{t}\left( L+L^{\dag }\right)
-j_{t}\left( \mathcal{L}_{G}X\right) \right) &=&0.
\end{eqnarray*}
Using the projective property of the conditional expectation $\pi _{t}((\pi
_{t}X))=\pi _{t}(X)$ and the assumption that $\alpha _{t}$ and $\beta _{t}$
lie in $\mathcal{Y}_{t}$, we find after a little algebra that 
\begin{eqnarray*}
\beta _{t} &=&\pi _{t}\left( XL+L^{\dag }X\right) -\pi _{t}\left( X\right)
\pi _{t}\left( L+L^{\dag }\right) , \\
\alpha _{t} &=&\pi _{t}\left( \mathcal{L}_{G}X\right) -\beta _{t}\pi
_{t}\left( L+L^{\dag }\right) ,
\end{eqnarray*}
so that the equation (\ref{filter ansatz}) reads as 
\begin{eqnarray}
d\pi _{t}(X) &=&\pi _{t}(\mathcal{L}_{G}X)dt  \label{eq:quant-filter-1} \\
&&+(\pi _{t}(XL+L^{\dag }X)-\pi _{t}(L+L^{\dag })\pi _{t}(X))dW(t),  \notag
\end{eqnarray}
where the \emph{innovations process} $W(t)$ is a Wiener process. It is
related to the measurement process $Y(t)$ by the equation 
\begin{equation}
dY(t)=\pi _{t}(L+L^{\dag })dt+dW(t)  \label{eq:outputquantrec-2}
\end{equation}
and has the interpretation as given the difference between the observed
change $dY(t)$ and the expected change\ $\pi _{t}(L+L^{\dag })dt$ in the
measured field immediately after time $t$. Note that the increment $dW(t)$ of the innovations process is independent of $\pi_s(X)$ for all 
$0 \leq s \leq t$.

It important to note that $Q(t)$ (equivalent to a Wiener process) and $W(t)$
(also a Wiener process) are distinct, and that $Q(t)$ is not in the
commutative observation subspace. Some care is needed in interpreting
equation (\ref{eq:quant-filter-1}) for the quantum filter. All of the terms
in this equation belong to the commutative subspace $\mathcal{Y}_{t}$, and
so (by the spectral theorem \cite{BouvanHJam07}) are statistically
equivalent to classical stochastic processes.

The stochastic master equation may be expressed in terms of the density
operator-valued stochastic process $\rho (t)$: 
\begin{equation*}
d\rho (t)=\mathcal{L}_{G}^{\star }\rho (t)dt+\mathcal{H}_{L}\rho(t)\,dW(t).
\end{equation*}
where we introduce 
\begin{equation}
\mathcal{H}_{L}\rho =L\rho + L^{\dag } \rho -\mathrm{tr}\left\{ \left( L+L^{\dag
}\right) \rho \right\} \, \rho .
\label{eq:calH}
\end{equation}
The increments $dW(t)$ can be generated independently of $\rho(s)$ for all
$0\leq s \leq t$ and the stochastic master equation above driven by the
generated increments can thus act as a simulated {\em quantum trajectory} 
of the state conditioned upon the measurement outcomes $\{y(s);\; 0 \leq s \leq t\}$.

\subsection{Photon Counting Case}

\label{sec:photon-count-basic}If instead we measure the number observable $%
Y(t)=U^{\dag }(t)\Lambda (t)U(t)=\Lambda _{out}(t)=\int_{0}^{t}b_{out}^{\dag
}(s)b_{out}(s)ds$ then the quantum filter is (see the survey paper \cite
{BouvanHJam07} for the derivation), 
\begin{equation*}
d\rho (t)=\mathcal{L}_{G}^{\star }\rho (t)dt+\mathcal{J}_{L}\rho
(t)\,dN(t)
\end{equation*}
where 
\begin{equation*}
\mathcal{J}_{L}\rho =\frac{L\rho L^{\dag }}{\mathrm{tr}\left\{ \rho L^{\dag
}L\right\} }-\rho ,
\end{equation*}
and the innovations process in this case is given by $dN(t)=dY-\mathrm{tr}%
\{\rho (t)L^{\dag }L\}dt$ and is a compensated Poisson process of
intensity $\mathrm{tr}\{\rho (t)L^{\dag }L\}$.

\subsection{Cascade Connections}

\label{sec:open-cascade} 
\begin{figure}[h]
\begin{center}
\includegraphics[scale=0.8]{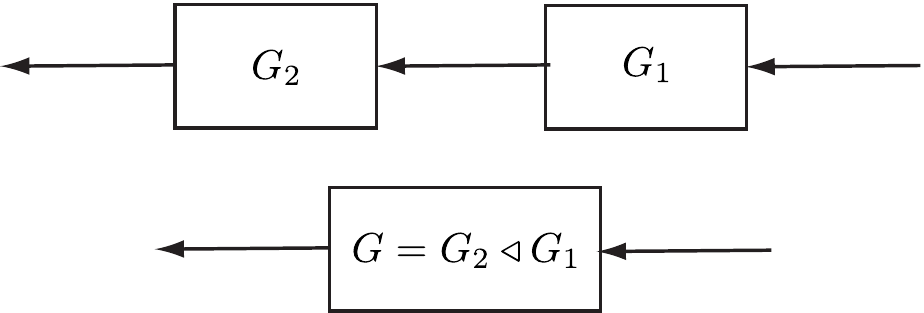} 
\end{center}
\caption{Two quantum systems cascaded, so that the output of system $G_{1}$
becomes the input of system $G_{2}$. Importantly the flow of information is
directional. In (b) the circuit has been simplified using quantum network
theory to an equivalent system $G=G_2 \triangleleft G_1$. This type of
network topology is called a cascade or series connection.}
\label{fig:series1}
\end{figure}

A simple quantum network may be formed by
connecting the output of one system to the input of another system, \cite
{Gar93,Car93,YanKim03_1,GouJam09b}. Fig. \ref{fig:series1} illustrates the
open quantum system $G=(S,L,H)$ equivalent to the cascade of systems $%
G_{1}=(S_{1},L_{1},H_{1})$ and $G_{2}=(S_{2},L_{2},H_{2})$. This equivalent
system can be described in terms of the \emph{series product} $%
G_{T}=G_{2}\triangleleft G_{1}$ \cite{GouJam09b}, defined by 
\begin{equation}
G_{2}\triangleleft G_{1}=(S_{2}S_{1},L_{2}+S_{2}L_{1},H_{1}+H_{2}+\mathrm{Im}%
\{ L_{2}^{\dagger }S_{2}L_{1}] \} ).  \label{series-product}
\end{equation}
Note in equation (\ref{series-product}) the order of the operators is
important. The series product provides the three parameters for the combined
or total open system $G$ in terms of the parameters for each of the systems $%
G_{1}$ and $G_{2}$.

\section{Single Photon Fields}

\label{sec:photon}

The master equation for a Markovian coupling of a system to a boson field in
a continuous-mode one or two photon state was first treated in ~\cite
{GheEllPelZol99}. In this section, we review the problem of determining the
associated filter (stochastic master equation) for an arbitrary system $%
G=(S,L,H)$ driven by a single photon field \cite{GJN11a}. In section~\ref
{sec:photon-mixed} we generalize the master equation and filter to include any combination of
single photon and vacuum as a probe field. The final section, section~\ref
{sec:photon-eg}, is an explicit example of the homodyne single photon
filtering equations for a two level atom.


\subsection{Continuous-Mode Single Photon States}

\label{sec:photon-state} 
There are many ways to generate single photon states \cite{single-photon-production}. One common technique for creating heralded
single photon states is by spontaneous parametric downconversion (SPDC). The
photons from such a process are inherently multimodal \cite{SPDC-mulitmode},
and spectral filtering is typically performed to get a single mode photon.

The creation operator for a photon with one-particle state $\xi $ is 
\begin{equation}
B^{\dag }\left( \xi \right) =\int_{0}^{\infty }\xi \left( t\right) dB^{\dag
}\left( t\right)
\end{equation}
normalized so that $\left\| \xi \right\| ^{2}=\int_{0}^{\infty }\left| \xi
\left( t\right) \right| ^{2}dt=1$. The single photon state is then defined
to be 
\begin{equation}
|1_{\xi }\rangle =B^{\dag }(\xi )|0\rangle .  \label{eq:xi-create}
\end{equation}
One may interpret this is the frequency domain as $|1_{\xi }\rangle
=\int_{-\infty }^{\infty }\hat{\xi}(\omega )\hat{b}^{\dag }(\omega
)|0\rangle $ where $\hat{\xi}$ is the Fourier transform of \ $\xi $ and\ $%
\hat{b}\left( \omega \right) $ the formal transform of the input process.
This representation is often referred to as the multimode, or
continuous-mode, single photon state, see for instance \cite[Sec. 6.3]
{Lou_book00}, \cite[Sec. 14.2]{Mil07}, \cite[Eq. (9)]{Mil08}.

Much of the calculations that follow will involve the identities 
\begin{eqnarray}
dB(t)|1_{\xi }\rangle &=&\xi (t)|0\rangle dt,  \notag \\
d\Lambda (t)|1_{\xi }\rangle &=&\xi (t)dB^{\dag }(t)|0\rangle ,
\label{eq:xi-annihilate-dB}
\end{eqnarray}
and this will be the origin of the departure of the master and filter
equations from the vacuum case.


\subsection{Single Photon Master Equation}

\label{sec:photon-master} 

Without loss of generality we fix the initial state of the system to be a
pure state $\rho _{0}=|\eta \rangle \langle \eta |$ and our aim is to obtain
a differential equation for the expectation 
\begin{equation*}
\varpi _{t}^{11}\left( X\right) =\langle \eta 1_{\xi }|j_{t}\left( X\right)
|\eta 1_{\xi }\rangle ,
\end{equation*}
for arbitrary system operator $X$. Starting from the Heisenberg-Langevin
equation as before, but now using the identities (\ref{eq:xi-annihilate-dB})
we find 
\begin{multline*}
\frac{d}{dt}\varpi _{t}^{11}\left( X\right) =\mathbb{E}_{11}[j_{t}(\mathcal{L%
}_{G}X))] \\
+\mathbb{E}_{01}[j_{t}(S^{\dag }[X,L])]\,\xi ^{\ast }(t)+\mathbb{E}%
_{10}[j_{t}([L^{\dag },X]S)]\,\xi (t) \\
+\mathbb{E}_{00}[j_{t}(S^{\dag }XS-X)]\,|\xi (t)|^{2} \\
=\varpi _{t}^{11}(\mathcal{L}_{G}X)+\varpi _{t}^{01}(S^{\dag }[X,L])\,\xi
^{\ast }(t) \\
+\varpi _{t}^{10}([L^{\dag },X]S)\,\xi (t)+\varpi _{t}^{00}(S^{\dag
}XS-X)\,|\xi (t)|^{2}
\end{multline*}
where 
\begin{eqnarray*}
\mathbb{E}_{jk}\left[ A\right] &=&\langle \eta \phi _{j}|A|\eta \phi
_{k}\rangle \\
\varpi _{t}^{jk}\left( X\right) &=&\mathbb{E}_{jk}[j_{t}\left( X\right) ]
\end{eqnarray*}
with 
\begin{equation*}
\phi _{j}=\left| 
\begin{array}{cc}
|0\rangle , & j =0; \\ 
|1_{\xi }\rangle , & j=1.
\end{array}
\right\}
\end{equation*}
Rather than finding a single master equation as in the vacuum case, we end
up with a system of equations 
\begin{align}
\dot{\varpi}_{t}^{11}(X) &=\varpi _{t}^{11}(\mathcal{L}X)+\varpi
_{t}^{01}(S^{\dag }[X,L])\xi ^{\ast }(t)  \notag \\
+&\varpi _{t}^{10}([L^{\dag },X]S)\xi (t)+\varpi _{t}^{00}(S^{\dag
}XS-X)|\xi (t)|^{2},  \notag \\
\dot{\varpi}_{t}^{10}(X) &=\varpi _{t}^{10}(\mathcal{L}X)+\varpi
_{t}^{00}(S^{\dag }[X,L])\xi ^{\ast }\!(t),  \notag 
\\
\dot{\varpi}_{t}^{01}(X) &=\varpi _{t}^{01}(\mathcal{L}X)+\varpi
_{t}^{00}([L^{\dag },X]S)\xi (t),  \notag \\
\dot{\varpi}_{t}^{00}(X) &=\varpi _{t}^{00}(\mathcal{L}X),
\label{eq:rho-dyn-a-00}
\end{align}
with initial conditions 
\begin{equation}
\varpi _{0}^{11}(X)=\varpi _{0}^{00}(X)=\langle \eta ,X\eta \rangle ,\ \
\varpi _{0}^{10}(X)=\varpi _{0}^{01}(X)=0.
\end{equation}
The main feature here is that the differential equation for expectations $%
\varpi ^{jk}$ depends on lower order $\varpi ^{jk}$, allowing us to solve
for $\varpi ^{11}$ inductively. Likewise, defining the traceclass operators $%
\varrho ^{jk}$ via 
\begin{equation}
\mathrm{tr}\left\{ \varrho^{jk}(t)^{\dag}X\right\} =\varpi _{t}^{jk}(X),
\end{equation}
we obtain a system of equations 
\begin{align}
\dot{\varrho}^{11}(t)& =\mathcal{L}^{\star }\varrho ^{11}(t)+\![S\rho
^{01}(t),L^{\dag }]\xi (t)+\![L,\varrho ^{10}(t)S^{\dag }]\xi ^{\ast }\!(t) 
\notag \\
& +(S\rho ^{00}(t)S^{\dag }-\varrho ^{00}(t))|\xi (t)|^{2},  \notag \\
\dot{\varrho}^{10}(t)& =\mathcal{L}^{\star }\varrho ^{10}(t)+[S\rho
^{00}(t),L^{\dag }]\xi (t),  \notag \\
\dot{\varrho}^{01}(t)& =\mathcal{L}^{\star }\varrho ^{01}(t)+[L,\varrho
^{00}(t)S^{\dag }]\xi ^{\ast }(t),  \notag \\
\dot{\varrho}^{00}(t)& =\mathcal{L}^{\star }\varrho ^{00}(t),
\label{eq:rho-dyn-00}
\end{align}
with 
\begin{equation*}
\varrho ^{11}(0)=\varrho ^{00}(0)=|\eta \rangle \langle \eta |,\ \ \varrho
^{10}(0)=\varrho ^{01}(0)=0.
\end{equation*}
Note $\varrho ^{jk}(t)^{\dag }=\varrho ^{kj}(t)$.


\subsection{An Input-Output Model of Single Photon Signal Generation}

\label{sec:photon-signal} 

\begin{figure}[h]
\begin{center}
\includegraphics[scale=0.6]{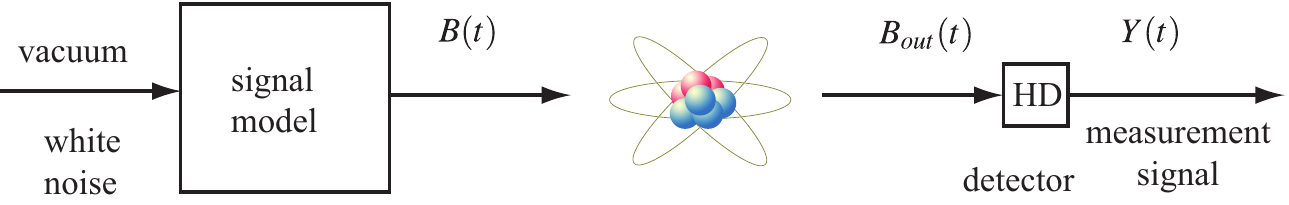} 
\end{center}
\caption{An ancilla system $M$ is used to model the effect of the single
photon state for $B(t)$ on the system $G$.}
\label{fig:system-signal}
\end{figure}

In section~\ref{sec:photon-filter} we will set up a general technique for
deriving the filtering equations for situations including the single photon
input field. It is possible to give an alternate derivation in this case
motivated by the idea of using a pre-interaction preparation where a vacuum
input is first passed through a fixed system in order to generate the one
photon field. Our motivation for considering such a scenario stems from statistical and engineering 
modelling where it is common practice to use `{signal generating
filters}' \cite{AM} driven by white noise to represent colored noise.
Analogously, in this section, we construct a quantum signal generating
filter $M=(S_{M},L_{M},H_{M})$. Cascading the single photon generating
filter $M$ with the quantum system $G$ we wish to probe, Figure \ref
{fig:system-signal}, we create an extended system. Because this extended
system $G_{T}=G\triangleleft M$ is driven by vacuum, the master equation and
quantum filter follow from the known vacuum case upon substitution of the
parameters for the cascade system (Section \ref{sec:photon-filter}).  We stress that
the signal generation model here (and in Section \ref{sec:cat-signal}  for the case of a system
driven by a superposition of continuous-mode coherent states) serves only as a convenient
theoretical mathematical device to derive the quantum filtering (or stochastic master)
equations. It is not suggested that single photons with a given wavepacket shape are to be generated in practice
with physical devices that implement this particular generator. 

The idea behind the signal generating filter $M$ is simple. We take the
filter to be a two level atom initially prepared in its excited state $%
|\uparrow \rangle $. The interaction with the vacuum input is taken to be 
\begin{equation}
(S_{M},L_{M},H_{M})=\left( I,\lambda \left( t\right) \sigma _{-},0\right) ,
\label{eq:photon-signal-model}
\end{equation}
which means that at some stage the atom decays into its ground state $%
|\downarrow \rangle $ creating a single photon in the output. The mechanism
for producing the single photon is therefore spontaneous emission due to the
coupling to the vacuum fluctuations. Here $\sigma _{-}$ is the lowering
operator from the upper state $|\uparrow \rangle $ to the ground state $%
|\downarrow \rangle $. The Schr\"{o}dinger equation for $|\psi _{t}\rangle
=V(t)|\uparrow \rangle \otimes |0\rangle $ then becomes $d|\psi _{t}\rangle =%
\left[ \lambda \left( t\right) \sigma _{-}dB_{t}^{\ast }-\frac{1}{2}\left|
\lambda \left( t\right) \right| ^{2}\sigma _{+}\sigma _{-}dt\right] \,|\psi
_{t}\rangle $, and it is an elementary calculation to see that this has the
exact solution 
\begin{equation}
|\psi _{t}\rangle =\sqrt{w\left( t\right) }|\uparrow \rangle \otimes
|0\rangle +|\downarrow \rangle \otimes B_{t}^{\ast }(\xi )|0\rangle
\end{equation}
where $B_{t}^{\ast }\left( \xi \right) =\int_{0}^{t}\xi _{s}dB_{s}^{\ast }$,
and (to preserve normalization) $w\left( t\right) =\int_{t}^{\infty }\left|
\xi \left( s\right) \right| ^{2}ds$ with the complex-valued function $\xi
\left( \cdot \right) $ related to $\lambda \left( \cdot \right) $ by 
\begin{equation}
\lambda \left( t\right) =\frac{1}{\sqrt{w\left( t\right) }}\xi \left(
t\right) .  \label{lambda from xi}
\end{equation}

Since $w\left(0 \right) =\left\| \xi \right\|
^{2}=1$, we therefore generate the limit state 
\begin{equation*}
|\psi _{\infty } \rangle =|\downarrow \rangle \otimes B^{\dag }\left( \xi \right)
|0\rangle \equiv |\downarrow \rangle \otimes |1_{\xi }\rangle .
\end{equation*}

Thus the generator model will output the desired single photon state $|1_\xi \rangle $ provided that  we choose the
(time-dependent) coupling strength $\lambda(t)$ according to (\ref{lambda from xi}).


\subsection{The Extended System}

\label{sec:extended-photon} 

We now define our extended system as the cascade system $G_{T}=G%
\triangleleft M$, as in Figure \ref{fig:system-signal}, where using the
cascade connection formalism from Section~\ref{sec:open-cascade} we have 
\begin{align}
G_{T}& =G\triangleleft M  \notag \\
& =\left( S,L+\frac{\xi (t)}{\sqrt{w(t)}}S\sigma _{-},H+\frac{\xi (t)}{\sqrt{%
w(t)}}\mathrm{Im}(L^{\dag }S\sigma _{-})\right) .  \label{eq:extend-photon}
\end{align}

Let us denote by $\tilde U(t)$ the unitary for the extended system driven by
vacuum for the parameters $G_T$ on the ancilla+system Hilbert space.
Specifying an initial state $| \uparrow \rangle \otimes | \eta \rangle
\otimes | 0 \rangle $, we consider the expectation 
\begin{equation}
\tilde \varpi_t(A \otimes X) = \mathbb{E}_{ \uparrow \eta 0}[ \tilde U^\dag
(t) (A \otimes X) \tilde U(t) ],  \label{eq:photon-extended-expect}
\end{equation}
(here $A$ is an ancilla operator, and $X$ is a system operator).

In order to be useful, the extended system $G_{T}$ (driven by vacuum) must
be capable of capturing expectations of $X(t)$, for arbitrary operator $X$
of the system $G$, at time $t$ as if it were driven by the single photon
field. That is, we must have 
\begin{equation}
\mathbb{E}_{\eta \xi }[X(t)]=\mathbb{E}_{\uparrow \eta 0}[\tilde{U}^{\dag
}(t)(I \otimes X \otimes I)\tilde{U}(t)],  \label{eq:photon-represent-1}
\end{equation}
that is we have the situation outlined in equation (\ref{eq:basic}) with 
\[ \rho_a = |\uparrow \rangle \langle \uparrow |, \quad
R(t) =I.
\]
We are required to  show that 
\begin{equation}
\mathbb{E}_{\eta \xi }[X(t)]=\mathbb{E}_{\uparrow \eta 0}[\tilde{U}^{\dag
}(t)(I\otimes X)\tilde{U}(t)]  \label{eq:photon-represent-1a}
\end{equation}
holds for any operator $X$ of the system $G$.

Our verification of (\ref{eq:photon-represent-1a}) is to compare the
differentials of both sides. Now the left hand side of (\ref
{eq:photon-represent-1a}) is just the single photon expectation $\varpi
_{t}^{11}(X)=\mathbb{E}_{11}[X(t)]=\mathbb{E}_{\eta \xi }[X(t)]$, whose
differential equation is determined from the system (\ref{eq:rho-dyn-a-00}).
The differential of the right hand side of (\ref{eq:photon-represent-1a})
may be found using the Lindblad superoperator $\mathcal{L}_{G_{T}}[A\otimes
X]\!$ for the extended system, which may be expressed in the form 
\begin{align*}
\mathcal{L}_{G_{T}}[A\otimes X]\!& =A\otimes \mathcal{L}_{G}X+(\mathcal{D}%
_{L_{M}}A)\otimes X \\
& +L_{M}^{\dag }A\otimes S^{\dag }[X,L]+AL_{M}\otimes \lbrack L^{\dag },X]S
\\
& +L_{M}^{\dag }AL_{M}\otimes (S^{\dag }XS-X),
\end{align*}
for any ancilla operator $A$ and system operator $X$. We first observe that

\begin{eqnarray}
&&\mathcal{D}_{L_{M}}(I)=0,\ \ \mathcal{D}_{L_{M}}(\sigma _{-})=-\frac{|\xi
(t)|^{2}}{2w(t)}\sigma _{-},  \notag \\
&&\mathcal{D}_{L_{M}}(\sigma _{+})=-\frac{|\xi (t)|^{2}}{2w(t)}\sigma _{+},\
\ \mathcal{D}_{L_{M}}(\sigma _{+}\sigma _{-})=-\frac{|\xi (t)|^{2}}{w(t)}%
\sigma _{+}\sigma _{-},  \notag
\end{eqnarray}
where $w(t)=\int_{t}^{\infty }|\xi (s)|^{2}ds$. Then we have 
\begin{eqnarray}
&&\frac{d}{dt}\mathbb{E}_{\uparrow \eta 0}[\tilde{U}^{\dag }(t)(I\otimes X)\tilde{U}%
(t)]=  \notag \\
&&\tilde{\varpi}_{t}^{11}(\mathcal{L}_{G}X)+\tilde{\varpi}_{t}^{01}(S^{\dag
}[X,L])\xi ^{\ast }(t)  \notag \\
&&+\tilde{\varpi}_{t}^{10}([L^{\dag },X]S)\xi (t)+\tilde{\varpi}%
_{t}^{00}(S^{\dag }XS-X)|\xi (t)|^{2},  \label{eq:hopefully-clearer-1}
\end{eqnarray}
where 
\begin{equation}
\tilde{\varpi}_{t}^{jk}(X)=\frac{\tilde{\varpi}_{t}(Q_{jk}\otimes X)}{%
w_{jk}(t)},  \label{eq:claim-mujk}
\end{equation}
with 
\begin{eqnarray*}
\left( Q_{jk}\right) &=&\left( 
\begin{array}{cc}
Q_{00} & Q_{01} \\ 
Q_{10} & Q_{11}
\end{array}
\right) =\left( 
\begin{array}{cc}
\sigma _{+}\sigma _{-} & \sigma _{+} \\ 
\sigma _{-} & I
\end{array}
\right) , \\
\left( w_{jk}\right) &=&\left( 
\begin{array}{cc}
w_{00} & w_{01} \\ 
w_{10} & w_{11}
\end{array}
\right) =\left( 
\begin{array}{cc}
w(t) & \sqrt{w(t)} \\ 
\sqrt{w(t)} & 1
\end{array}
\right) .
\end{eqnarray*}
Notice that equation (\ref{eq:hopefully-clearer-1}) for $\tilde{\varpi}%
_{t}^{11}(X)$ has the same form as the $\varpi _{t}^{11}(X)$ equation in (%
\ref{eq:rho-dyn-a-00}). In general, the equations for $\tilde{\varpi}%
_{t}^{jk}(X)$ have the same form as equations (\ref{eq:rho-dyn-a-00}) for $%
\varpi _{t}^{jk}(X)$. Since at time $t=0$ we have $\tilde{\varpi}%
_{0}^{jk}(X)=\varpi _{0}^{jk}(X)$, it follows that $\tilde{\varpi}%
_{t}^{jk}(X)=\varpi _{t}^{jk}(X)$ for all $t$. This establishes the identity
(\ref{eq:photon-represent-1a}).


\subsection{Single Photon Stochastic Master Equation (Filter) for Quadrature
Phase Measurements}

\label{sec:photon-filter} 
In this section we explain how the quantum filter for the conditional
expectation 
\[
\pi_t^{11}(X) = \mathbb{E}_{\eta\xi}[ X(t) \vert Y(s), 0 \leq s
\leq t ] 
\]
for the system $G$ driven by a single photon field may now be
obtained from the quantum filter for the conditional expectation 
\begin{equation}
\tilde \pi_t(A \otimes X) = \mathbb{E}_{\uparrow \eta 0}[ \tilde U^\dag(t) (A
\otimes X) \tilde U(t) \vert I \otimes Y(s), 0 \leq s \leq t]
\label{eq:photon-extended-ce-def}
\end{equation}
for the extended system $G_{T}=G \triangleleft M$ driven by vacuum.

Indeed, we have 
\begin{eqnarray}
d\tilde \pi_t(A \otimes X) &=& \tilde \pi_t( \mathcal{L}_{G_T}(A\otimes X) )
dt  \notag \\
&& + (\tilde \pi_t( A\otimes XL_T + L_T^\dag A\otimes X)  \notag \\
&& - \tilde \pi_t(L_T+L_T^\dag) \tilde \pi_t(A\otimes X)) d W(t),
\label{eq:photon-extended-filter}
\end{eqnarray}
where $dW(t) = dY(t) - \tilde \pi_t(L_T+L_T^\dag)dt$. If we define 
\begin{equation}
\pi^{jk}_t(X) = \frac{\tilde\pi_t( Q_{jk} \otimes X) }{ w_{jk}(t) },
\label{eq:pi-jk-photon}
\end{equation}
where $Q_{jk}$ and $w_{jk}(t)$ were defined in the previous section, we
obtain the coupled system of nonlinear stochastic differential equations 
\begin{widetext}
\begin{align}
d\pi^{11}_t (X) =&  \bigl \{ \pi^{11}_t(\mathcal{L}X) + \pi^{01}_t( S^\dag [X,L] )
\xi^\ast(t)  + \pi^{10}_t( [L^\dag, X] S ) \xi(t)  + \pi^{00}_t( S^\dag X S - X) \vert \xi(t) \vert^2\bigr \}dt  \notag \\
& + \big \{ \pi^{11}_t( XL + L^\dag X) + \pi^{01}_t(S^\dag X) \xi^\ast(t) +
\pi^{10}_t( XS) \xi(t)   - \pi^{11}_t(X) K_t \big\}dW(t),  \notag
 \\
d\pi^{10}_t (X) =& \bigl \{ \pi^{10}_t(\mathcal{L}X) + \pi^{00}_t( S^\dag [X, L]
) \xi^\ast(t) \bigr \}dt  +  \big \{( \pi^{10}_t( XL + L^\dag X) + \pi^{00}_t(S^\dag X) \xi^\ast(t)  - \pi^{10}_t(X)  K_t
  \big \}dW(t),  
\notag \\
d\pi^{01}_t (X) =& \bigl \{ \pi^{01}_t(\mathcal{L}X) + \pi^{00}_t( S^\dag [X, L]
) \xi^\ast(t) \bigr \}dt  +  \big \{( \pi^{01}_t( XL + L^\dag X) + \pi^{00}_t(S^\dag X) \xi^\ast(t)  - \pi^{01}_t(X)  K_t
  \big \}dW(t),  
\notag \\
d\pi^{00}_t (X) =& \pi^{00}_t(\mathcal{L}X) dt + \big \{ \pi^{00}_t( XL +
L^\dag X)   - \pi^{00}_t(X)  K_t \big \}dW(t) . 
 \label{eq:pi-dyn-a-00}
\end{align}
\end{widetext}
Here, 
\begin{equation}
K_t = \pi^{11}_t(L+L^\dag) + \pi^{01}_t(S) \xi(t) + \pi^{10}_t(S^\dag)
\xi^\ast(t)
\end{equation}
and the innovations process $W(t)$ (given above) may be expressed as 
\begin{equation}
dW(t) = dY(t) - K_t dt .  \label{eq:innovation-11}
\end{equation}
We have $\pi^{01}_t(X) = \pi^{10}_t(X^\dag)^\dag$, and the initial
conditions are $\pi^{11}_0(X)= \pi^{00}_0(X)= \langle \eta, X \eta \rangle,
\ \ \pi^{10}_0(X)= \pi^{01}_0(X)=0. $

In order to see that the single photon quantum filter is given by the system
of coupled equations (\ref{eq:pi-dyn-a-00}), we must show that the
conditional expectation for the system driven by the single photon field is
given by 
\begin{eqnarray}
\pi _{t}(X) &=&\mathbb{E}_{\eta \xi }[X(t)|Y(s),\ 0\leq s\leq t]  \notag \\
&=&\mathbb{E}_{\uparrow \eta 0}[\tilde{U}^{\dag }(t)(A\otimes X)\tilde{U}%
(t)|I\otimes Y(s),0\leq s\leq t]  \notag \\
&=&\pi _{t}^{11}(X).  \label{eq:photon-pi11-def}
\end{eqnarray}
To obtain the filter, we again apply the characteristic function technique,
setting $C=C_{g}(t)$ as before with $dc_{g}(t)=g(t)c_{g}(t)dY(t)$. We need
to verify that 
\begin{equation}
\mathbb{E}_{\eta 0}[j_{t}(X)c_{g}(t)]=\mathbb{E}_{\eta 0}[\pi
_{t}(X)c_{g}(t)]  \label{eq:g-photon-goal-1}
\end{equation}
For the extended system we have 
\begin{equation}
\mathbb{E}_{\uparrow \eta 0}[\tilde{U}^{\dag }(t)(A\otimes X)\tilde{U}(t)c_{g}(t)]=%
\mathbb{E}_{\eta 0}[\tilde{\pi}_{t}(A\otimes X)c_{g}(t)],
\label{eq:g-photon-extend}
\end{equation}
for all functions $g$, arbitrary ancilla, and system operators $A$ and $X$
respectively. Hence (\ref{eq:g-photon-goal-1}) will follow provided we can
show that 
\begin{equation}
\mathbb{E}_{jk}[X(t)c_{g}(t)]=\frac{\mathbb{E}_{e\eta 0}[\tilde{U}^{\dag
}(t)(Q_{jk}\otimes X)\tilde{U}(t)c_{g}(t)]}{w_{jk}(t)}.
\label{eq:g-photon-goal-2}
\end{equation}
However, equation (\ref{eq:g-photon-goal-2}) may be verified in exactly the
same way we proved that $\tilde{\varpi}_{t}^{jk}(X)=\varpi _{t}^{jk}(X)$ in
the previous section, that is, by comparing the differentials of both sides
of (\ref{eq:g-photon-goal-2}). The details of this calculation are omitted.

Now, write $\pi _{t}^{jk}(X)=\mathrm{tr}((${$\varrho $}$^{jk}(t))^{\dag }X)$%
. Then from the differential equations for $\pi_t^{jk}(X)$ and the definition 
{$\varrho $}$^{jk}(t)$ we immediate get the differential equations for the
evolution of {$\varrho $}$^{jk}(t)$, as follows:

\begin{widetext}
\begin{eqnarray}
d   \rho^{11} (t) &=& \bigl \{\mathcal{L}^\star  \rho^{11} (t) + [
S \rho^{01}(t), L^\dag] \xi(t)   + [ L ,  \rho^{10}(t) S^\dag  ]
\xi^\ast(t) + (S  \rho^{00}(t) S^\dag - \rho^{00}(t)) \vert \xi(t) \vert^2 \big \} dt \notag   \\
&& \notag + \big \{ L  \rho^{11}(t) +   \rho^{11}(t) L^\dag +   \rho^{10}(t)
S^\dag \xi^\ast(t) + S   \rho^{01}(t) \xi(t)  -  K_t    \rho^{11}(t) \big \}  dW(t),  \notag  \\
d   \rho^{10} (t) &=& \big \{ \mathcal{L}^\star  \rho^{10} (t)  + [ S 
\rho^{00}(t), L^\dag ] \xi(t) \big \}dt   +\big \{ L  \rho^{10}(t) +   
\rho^{10}(t) L^\dag + S   \rho^{00}(t)
\xi(t)  -  K_t    \rho^{10}(t) \big \}  dW(t),  \notag \\
d   \rho^{01}(t)&=& \big \{ \mathcal{L}^\star   \rho^{01}(t)  + [  L,
 \rho^{00} (t) S^\dag ] \xi^\ast(t) \big \}dt  + \big \{ L  \rho^{01}(t) +   
\rho^{01}(t) L^\dag +   \rho^{00}(t)
S^\dag \xi^\ast (t)   -  K_t    \rho^{01}(t) \big \}  dW(t),  \notag \\
d   \rho^{00} (t) &=& \mathcal{L}^\star   \rho^{00}(t)  dt + \big \{ L 
\rho^{00}(t) +   \rho^{00}(t) L^\dag   -  K_t    \rho^{00}(t) \big \}  dW(t), \label{eq:hat-rho-dyn-00}
\end{eqnarray}
\end{widetext}
where 
\begin{align*}
K_{t} &\equiv  \mathrm{tr}\{{{(L+L^{\dag })\rho ^{11}(t)\}}}  \\
&+\mathrm{tr}\{S\rho {{^{01}(t)\}}}\xi (t)+\mathrm{tr}\{{{S^{\dag
}\rho ^{10}(t)\}}}\xi ^{\ast }(t),
\end{align*}
with the initial condition
\begin{align*}
\rho ^{11}(0)&=\rho ^{00}(0)=|\eta \rangle \langle \eta |,\;
\rho ^{10}(0)=\rho ^{01}(0)= 0.
\end{align*}

\subsection{Single Photon Stochastic Master Equation (Filter) for Photon Counting Measurements}

\label{sec:photon-counting} 
In this section we briefly derive the filtering equations for photon
counting measurements. The quantum filter for the photon counting case is
given by the system of equations
\begin{widetext}
\begin{eqnarray*}
d\pi^{11}_t (X)&=& \{ \pi^{11}_t(\mathcal{L}X) + \pi^{01}_t( S ^\dag  [X,L] )
\xi^\ast(t) + \pi^{10}_t( [L ^\dag , X] S ) \xi(t)   + \pi^{00}_t( S ^\dag  X S - X) \vert \xi(t) \vert^2 \} dt  \notag \\
&& 
+ \biggl\{ \nu_t^{-1} \left( \pi^{11}_t(L ^\dag  XL) + \pi^{01}_t(S ^\dag  X L) \xi^\ast(t) + \pi^{10}_t( L ^\dag  X S) \xi(t) + \pi^{00}_t( S ^\dag  XS) \vert \xi(t) \vert^2 
\right) - \pi^{11}_t(X)
\biggr\} dN(t), \\
{d\pi^{10}_t (X)}&=& \{\pi^{10}_t(\mathcal{L}X) + \pi^{00}_t( S ^\dag  [X,L] )\xi^\ast(t) \}dt   
+ \biggl\{  \nu_t^{-1} \left( \pi^{10}_t(L ^\dag  XL) + \pi^{00}_t(S ^\dag  X L) \xi^\ast(t)\right)
- \pi^{10}_t(X)\biggr\}  dN(t),\\
d\pi^{01}_t (X) &=& \{\pi^{01}_t(\mathcal{L}X) + \pi^{00}_t(  [L ^\dag ,X] S )\xi(t) \}dt   
+ \bigg \{ \nu_t^{-1} \left( \pi^{01}_t(L ^\dag  XL) + \pi^{00}_t(L ^\dag  X S) \xi (t)  \right) 
 - \pi^{01}_t(X)
\bigg \}  dN(t),\\
d\pi^{00}_t (X)&=& \pi^{00}_t(\mathcal{L}X)  dt   
+ \biggl\{  \nu_t^{-1} \bigl(\pi^{00}_t(L ^\dag  XL) \bigr) - \pi^{01}_t(X)
\biggr\}  dN(t),
\end{eqnarray*}
or in the \sch-picture
\begin{eqnarray}
d\rho^{11}(t)&=&\bigl \{\mathcal{L}^\star  \rho^{11} (t) + [
S \rho^{01}(t), L^\dag] \xi(t)   + [ L ,  \rho^{10}(t) S^\dag  ]
\xi^\ast(t) + (S  \rho^{00}(t) S^\dag - \rho^{00}(t)) \vert \xi(t) \vert^2 \big \} dt \notag   \\
&& 
+ \biggl\{ \nu_t^{-1} \left( L\rho^{11}(t) L ^\dag  + L\rho^{10}(t)S ^\dag   \xi^\ast(t) + S\rho^{10}(t) L ^\dag \xi(t) + S\rho^{00}(t) S ^\dag  \vert \xi(t) \vert^2 
\right) - \rho^{11}(t)
\biggr\} dN(t), \notag\\
d   \rho^{10} (t) &=& \big \{ \mathcal{L}^\star  \rho^{10} (t)  + [ S \rho^{00}(t), L^\dag ] \xi(t) \big \}dt
+ \biggl\{  \nu_t^{-1} \left( L\rho^{10}(t)L ^\dag + S\rho^{00}(t)L ^\dag  \xi(t)\right)
- \rho^{10}(t)\biggr\}  dN(t),\notag\\
d   \rho^{01} (t) &=& \big \{ \mathcal{L}^\star  \rho^{01} (t)  + [ L,\rho^{00}(t) S^\dag ] \xi^*\!(t) \big \}dt
 + \bigg \{ \nu_t^{-1} \left(  L\rho^{01}(t)L ^\dag  + L\rho^{00}(t)S ^\dag  \xi^*(t)  \right) 
 - \rho^{01}(t)
\bigg \}  dN(t),\notag\\
d \rho^{00}(t)&=&  \mathcal{L}^\star   \rho^{00}(t)  dt 
  + \biggl\{  \nu_t^{-1} \bigl( L\rho^{00}(t)L ^\dag  \bigr) - \rho^{00}(t)
\biggr\}  dN(t), \label{eq:counting}
\end{eqnarray}

\end{widetext}
where 
\begin{align*}
\nu_t &= \pi^{11}_t(L ^\dag  L) + \pi^{01}_t(S ^\dag   L) \xi^\ast(t)  \\ 
&+ \pi^{10}_t( L ^\dag   S) \xi(t) + \pi^{00}_t( I) \vert \xi(t) \vert^2,\\
 &= \tr{\rho^{11}(t)L ^\dag  L} + \tr{\rho^{10}(t) S ^\dag   L} \xi^\ast(t)  \\ 
& +\tr{ \rho^{01}(t) L ^\dag   S} \xi(t) + \tr{\rho^{00}(t)I} \vert \xi(t) \vert^2,
\end{align*}
and the innovations process $N(t)$  is given by 
\[ 
dN(t)= dY(t) - \nu_t  dt .
\]


\subsection{Combination of One Photon and Vacuum States}

\label{sec:photon-mixed} 

In this section we take the state of the field to be in a state defined by
the density operator 
\begin{equation}
\rho _{\mathrm{field}}=\sum_{jk}\gamma _{kj}|\phi _{j}\rangle \langle \phi _{k}|
\label{eq:photon-vac-density}
\end{equation}
where we use the notation introduced above for the photon $|\phi _{1}\rangle
=|1_{\xi }\rangle $ and vacuum $|\phi _{0}\rangle =|0\rangle $ states. The
coefficients $\gamma _{jk}$ must of course satisfy the condition that the $2\times 2$
complex matrix 
\begin{equation}
\rho _{a}=\sum_{jk}\gamma _{kj}|j\rangle \langle k|=\left( 
\begin{array}{cc}
\gamma _{11} & \gamma _{10} \\ 
\gamma _{01} & \gamma _{00}
\end{array}
\right)  \label{eq:photon-vac-density-ancilla}
\end{equation}
is a density matrix, i.e. $\rho _{a}\geq 0$, $\mathrm{Tr}[\rho _{a}]=1$. By
choosing the coefficients $\gamma _{jk}$ appropriately we can model an input
field that is any combination of single photon and vacuum. For example: the
single photon field is given by $\gamma _{11}=1$ and all other coefficients
are zero, a superposition like $\left| {\psi }\right\rangle _{f}=\alpha
_{1}\left| {1_{\xi }}\right\rangle +\alpha _{0}\left| {0}\right\rangle $ is
obtained by setting $\gamma _{11}=|\alpha _{1}|^{2},\,\gamma _{10}=\alpha
_{1}\alpha _{0}^{\ast },\,\gamma _{01}=\alpha _{0}\alpha _{1}^{\ast
},\,\gamma _{00}=|\alpha _{0}|^{2}$; and a simple combination is $\rho _{\mathrm{field}}=\eta
\left| {1}\right\rangle \!\left\langle {1}\right| +(1-\eta )\left| {0}%
\right\rangle \!\left\langle {0}\right| $ where $\gamma _{11}=\eta ,\,\gamma
_{00}=1-\eta $ and $\gamma _{10}=\gamma _{01}=0$.

\subsubsection{The Master Equation}
\begin{center}
\begin{figure}[h]
\includegraphics[width= \linewidth]{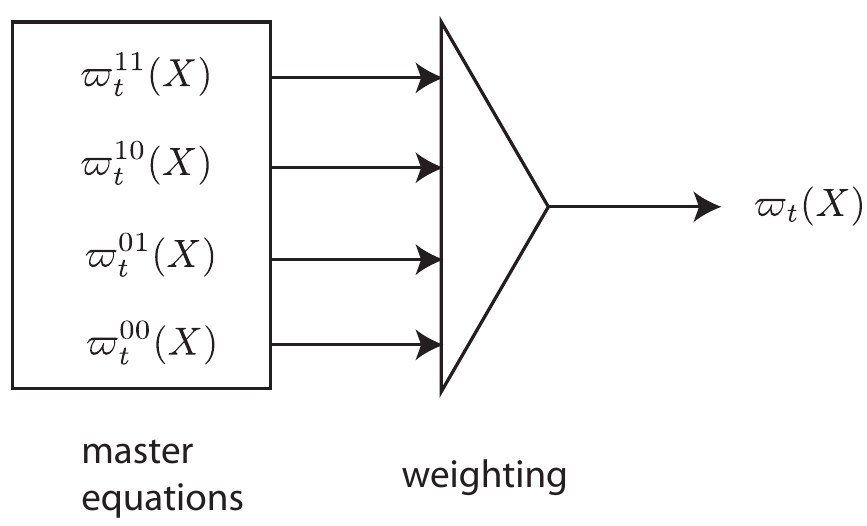}
\caption{The expectation $\protect\varpi _{t}(X)=\langle X(t)\rangle $ of
the system operator $X(t)=j_t (X)$ when the system and field are initialized in the
state $|\protect\eta \rangle \langle \protect\eta |\otimes \protect\rho _{\mathrm{field}}$
may be calculated by weighting the solutions $\protect\varpi _{t}^{jk}(X)$
from the single photon master equations (\ref{eq:rho-dyn-a-00}).}
\label{fig:mixed-master1}
\end{figure}
\end{center}

The expectation $\varpi _{t}(X)=\langle X(t)\rangle $ of the system operator 
$X(t)$ when the system and field are initialized in the state $|\eta \rangle
\langle \eta |\otimes \rho _{\mathrm{field}}$ is given by 
\begin{eqnarray}
\varpi _{t}(X)=\mathbb{E}_{\eta \rho _{\mathrm{field}}}[X(t)] &=&\sum_{jk}\gamma _{jk}%
\mathbb{E}_{jk}[X(t)]  \notag \\
&=&\sum_{jk}\gamma _{jk}\varpi _{t}^{jk}(X),  \label{eq:photon-master-mixed}
\end{eqnarray}
where $\varpi _{t}^{jk}(X)$ are defined in section \ref{sec:photon-master}.
While there is no differential equation for $\varpi _{t}(X)$, it can be
computed from the weighted sum (\ref{eq:photon-master-mixed}), Figure \ref
{fig:mixed-master1}. From equation (\ref{eq:photon-master-mixed}) we see
that the density operator for the expectation $\varpi _{t}(X)=\mathrm{tr}%
\left\{ \varrho (t)X\right\} $ is given by 
\begin{equation}
\varrho (t)=\sum_{jk}\gamma _{kj}\varrho ^{jk}(t),
\end{equation}
where the $\varrho ^{jk}(t)$ are the density operators introduced in {section 
\ref{sec:photon-master}}.

\subsubsection{The Stochastic Master Equation}
Turning now to the problem of determining the filter, we again make use of
the cascade extended system from Section \ref{sec:extended-photon}. Now we
have 
\begin{equation}
\varpi _{t}^{jk}(X)=\frac{\tilde{\varpi}_{t}(Q_{jk}\otimes X)}{w_{jk}(t)},
\label{eq:claim-mujk-ok}
\end{equation}
and so if we define the matrix 
\begin{equation}
R(t)=\sum_{jk}\frac{\gamma _{jk}}{w_{jk}(t)}Q_{jk},
\end{equation}
where $w_{jk}(t)$ and $Q_{jk}$ are as defined in Section \ref
{sec:extended-photon}, we have (using (\ref{eq:claim-mujk-ok}), (\ref
{eq:photon-master-mixed}) and (\ref{eq:photon-extended-expect})) 
\begin{equation}
\varpi _{t}(X)=\tilde{\varpi}_{t}(R(t)\otimes X).
\end{equation}
Note that the definition (\ref{eq:photon-extended-expect}) of $\tilde{\varpi}%
_{t}(R(t)\otimes X)$ involves the ancilla system initialized in the excited
state $|e\rangle =|\uparrow \rangle $.

The conditional expectation 
\begin{equation}
\pi _{t}(X)=\mathbb{E}_{\eta \rho _{\mathrm{field}}}[X(t)\,|\,Y(s),0\leq s\leq t]
\end{equation}
corresponding to the field in the state $\rho _{\mathrm{field}}$ is related to the
conditional expectation $\tilde{\pi}_{t}(A\otimes X)$ for the extended
system (see (\ref{eq:photon-extended-ce-def})) by the Bayes relation 
\begin{equation}
\pi _{t}(X)=\frac{\tilde{\pi}_{t}(R(t)\otimes X)}{\tilde{\pi}%
_{t}(R(t)\otimes I)}.  \label{eq:bayes-R}
\end{equation}
Division by the denominator in (\ref{eq:bayes-R}) is needed to ensure
the normalization $\pi _{t}(I)=1$. To prove (\ref{eq:bayes-R}), we need to show
that $\tilde{\pi}_{t}(R(t)\otimes X)=\tilde{\pi}_{t}(R(t)\otimes I)\pi
_{t}(X)$, or equivalently 
\begin{equation*}
\mathbb{E}_{\uparrow \eta 0}[\tilde{\pi}_{t}(R(t)\otimes X)c_{g}(t)]=\mathbb{E}%
_{\uparrow \eta 0}[\tilde{\pi}_{t}(R(t)\otimes I)\pi _{t}(X)c_{g}(t)]
\end{equation*}
for all choice of characteristic functions $c_{g}\left( t\right) $. However, 
$ \mathbb{E}_{\uparrow \eta 0}[\tilde{\pi}_{t}(R(t)\otimes X)c_{g}(t)]$ equals $\mathbb{E}%
_{\uparrow \eta 0}[\tilde{U}^{\dag }(t)(R(t)\otimes X\otimes I)\tilde{U}(t)c_{g}(t)]$,
but by the extended system representation this is just $\mathbb{E}_{\eta
\rho _{\mathrm{field}}}[X(t)c_{g}(t)]$ which in turn equals $\mathbb{E}_{\eta \rho
_{f}}[\pi _{t}(X)c_{g}(t)]\equiv \mathbb{E}_{\uparrow \eta 0}[\tilde{U}^{\dag
}(t)(R(t)\otimes I)\tilde{U}(t)\pi _{t}(X)c_{g}(t)]$ which establishes the
Bayes relation (\ref{eq:bayes-R}).

Since 
\begin{equation}
\tilde \pi_t(R(t) \otimes X) = \sum_{jk} \gamma_{jk} \pi^{jk}_t(X),
\end{equation}
where $\pi^{jk}_t(X)$ is defined by (\ref{eq:pi-jk-photon}), the desired
conditional expectation may be expressed as 
\begin{eqnarray}
\pi_t(X) = \frac{ \sum_{j,k} \gamma_{jk} \pi^{jk}_t(X) }{ \sum_{j,k}
\gamma_{jk} \pi^{jk}_t(I) }.  \label{eq:photon-mixed-filter}
\end{eqnarray}

\begin{center}
\begin{figure}[h]
\includegraphics[width= \linewidth]{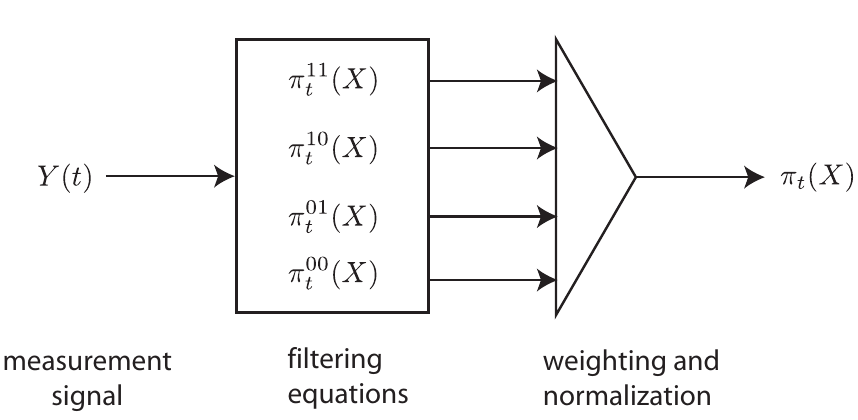}
\caption{Relationship between the measured signal and the filtered estimate.
The differential equations (\ref{eq:pi-dyn-a-00}) must be integrated to
compute $\protect\pi_{t}^{jk}(X)$. Then depending on the state of the input
field, the probe, these estimates must be weighted by the appropriate
coefficients and normalized as specified in (\ref{eq:photon-mixed-filter})
to produce the desired conditional expectation $\protect\pi_{t}(X)$.}
\label{fig:photon-filter-mixed}
\end{figure}
\end{center}

Again, there is no differential equation for $\pi _{t}(X)$; instead it is
computed from a normalized weighted sum, (\ref{eq:photon-mixed-filter}), and
the filtering equations (\ref{eq:pi-dyn-a-00}), Figure \ref
{fig:photon-filter-mixed}. The corresponding conditional density operator is
given by 
\begin{equation}
\rho (t)=\frac{\sum_{jk}\gamma _{kj}\rho ^{jk}(t)}{\sum_{jk}\gamma
_{kj}\mathrm{tr}\left\{ \rho ^{jk}(t)\right\} },
\label{eq:photon-cdl-density}
\end{equation}
where the conditional quantities $\rho ^{jk}(t)$ may be computed from the
single photon filtering equations (\ref{eq:hat-rho-dyn-00}).

These expressions allow filtering on any combination of a single photon and
a vacuum state. One notable case it that of simple combination of one photon and vacuum (%
$\rho _{\mathrm{probe}}=p\left| {1}\right\rangle \!\left\langle {1}\right|
+(1-p)\left| {0}\right\rangle \!\left\langle {0}\right| $) which is an
experimentally accurate model for the output of the SPDC process \cite
{LvoAicBen01}.


\subsection{Illustrative Example of Single Photon Master and Filtering Equations}

\label{sec:photon-eg} 

Here apply the filtering method derived above to the problem of exciting a two level atom, in free space, with a continuous mode single photon. This problem has received much attention recently \cite{StoAlbLeu07,WanSheSca10,StoAlbLeu10,RepSheFan10}. Until now it has only been possible to calculate ensemble averaged quantities. Here we show the individual trajectories associated with a particular experimental run.

This problem can be parametrized in our model as follows. We take the coupling operator to be $L=\sqrt{\kappa }\sigma _{-}$, the internal dynamics of the atom are specified by the Hamiltonian $H=0$ and there is no scattering i.e. $S=I$. Here $\kappa >0$ is the coupling rate (often referred to as the measurement strength) and is chosen to be $\kappa = 1$. The atom is take to be in the ground state initially $\op{g}{g}$, then a single photon in the wavepacket $\xi(t)$ interacts with the atom. We take the wavepacket  to be a Gaussian parametrized as
\begin{align}\label{Eq::gau_xi}
\xi_{\rm gau}(t)=\left( \frac{\Omega^2}{2 \pi} \right)^{1/4} \!\exp{\left[ -\frac{\Omega^2}{4} (t - t_c )^ 2\right] } ,
\end{align}
where $t_c$ specifies the peak arrival time and $\Omega$ is the frequency bandwidth of the pluse. 

\begin{center}
\begin{figure}[h]
\includegraphics[width= \linewidth]{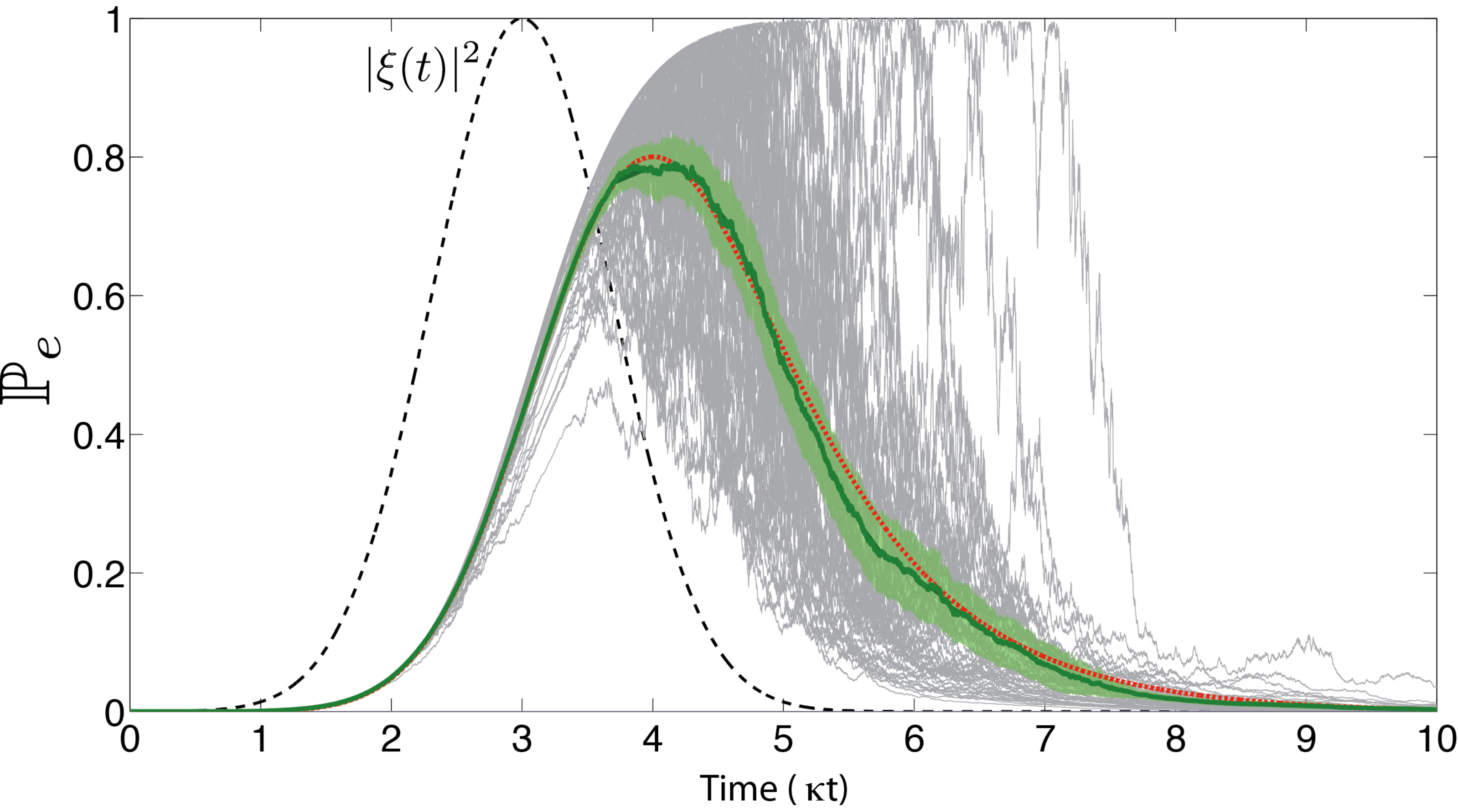}
\caption{The excited state population, $\mathbb{P}_e$, of a two-level atom interacting with one photon in a Gaussian wavepacket. The dashed line is the Gaussian wavepacket $|\xi(t)|^2$ with bandwidth $\Omega = 1.46 \kappa$. The dotted (red) line is $\mathbb{P}_e$ as calculated by the master equation. The grey lines are the individual trajectories $\mathbb{P}_e^c$. The solid line is the ensemble average of sixty four trajectories plotted with error bars (the shaded light green region).}
\label{fig:exitation}
\end{figure}
\end{center}

Now we wish to calculate the excited state population of the two level atom as a function of time. Other studies have only been able to calculate the master equation evolution of the atomic state \cite{StoAlbLeu07,WanSheSca10,StoAlbLeu10,RepSheFan10}. In our formalism this corresponds to propagating the master equations and taking the expectation 
\begin{align}\label{eq:Pe_me}
\mathbb{P}_e(t) & = \tr { \varrho^{11}(t) \op{e}{e} },
\end{align} 
where $\varrho^{11}(t) $ is the solution to \erf{eq:rho-dyn-00}. In Fig. \ref{fig:exitation} \erf{eq:Pe_me} is plotted, the dotted line (red), as a function of time for a two level atom interacting with a gaussian pulse. We choose $\Omega = 1.46 \kappa$ which is known to be optimal for excitation via a single photon in a Gaussian pulse   \cite{StoAlbLeu07,WanSheSca10,StoAlbLeu10}. Our numerics agree with the prior results that $\max_t \mathbb{P}_e(t)\approx 0.8$ \cite{StoAlbLeu07,WanSheSca10,StoAlbLeu10}.

However, in our formalism we can also calculate the conditional state of the system using the quantum filtering equations derived above. The conditional excited state population is denoted by 
\begin{align}\label{eq:Pe_sme}
\mathbb{P}_e^c(t) & = \tr { \rho^{11}(t) \op{e}{e} },
\end{align} 
where $\rho^{11}(t) $ is the solution to the filtering equations \erf{eq:hat-rho-dyn-00} or \erf{eq:counting} for homodyne or photon counting measurements respectively. In what follows we will focus on the homodyne measurement filtering equations i.e. \erf{eq:hat-rho-dyn-00}.

In Fig. \ref{fig:exitation}, 64 different trajectories given by \erf{eq:Pe_sme} are plotted as grey lines. For this particular bandwidth there is very little spread in the trajectories for  $t<3$. After the bulk of the wavepacket has passed, at $t=4$, many of the trajectories start to decay, as evidenced by the many grey lines below $\mathbb{P}_e^c = 0.5$ for $t>4$. Nevertheless there are a number of trajectories which continue to rise towards $\mathbb{P}_e^c = 1$ for $t>4$. This means in a particular run of an experiment the atom may become fully excited. Such behavior can not be seen through the master equation approach of Refs.~\cite{StoAlbLeu07,WanSheSca10,StoAlbLeu10,RepSheFan10}. 

It is possible to confirm the consistency of the trajectories with the master equation solution by calculating a numerical average of the trajectories. We plot the ensemble average of the trajectories as the solid line in Fig.~\ref{fig:exitation} with error bars smeared around this line. The numerically calculated ensemble average agrees with the master equation behavior given that a small ensemble was used to calculate this mean value.

\section{Superposition of Coherent Field States}

\label{sec:cat} 

In this section we turn to the problem of determining the master equation
and the quantum filter for systems drive by a boson field whose state is a
superposition of  continuous-mode coherent states. In section~\ref
{sec:cat-state} we describe continuous-mode coherent states and
superpositions of them, as well as the action of the quantum noises on such
states. Section \ref{sec:cat-master} is devoted to the derivation of the
master equation for superpositions of coherent states. In section~\ref
{sec:cat-signal} we develop a cascaded system signal model. This model
allows us to use the methodology from Section \ref{sec:photon}, with
appropriate changes due to the nature of the superposition of coherent
states, to derive the filtering equations in section~\ref{sec:cat-filter}.
Then we give the filter for the case of photon counting in section~\ref
{sec:super-counting} and generalize to mixed input states.


\subsection{Superpositions and Combinations of Coherent States}

\label{sec:cat-state} 

Typically single mode coherent states of a field are denoted by $\left| {%
\alpha }\right\rangle $. In this paper we shall often refer to
a superposition of continuous-mode coherent states as a (continuous-mode)
cat-state \cite{Lou_book00,GarChi_book08}. Formally, the superpositions of
continuous-mode coherent states is given by 
\begin{equation}
|\psi \rangle =\sum_{j=1}^{n}s_{j}|\alpha _{j}\rangle ,
\label{eq:super-state}
\end{equation}
where $|\alpha _{j}\rangle $ are coherent states, determined by functions $%
\alpha _{j}(t)$ with $\alpha _{j}\neq \alpha _{k}$ if $j\neq k$. The
superposition weights $s_{j}$ are complex numbers such that $\langle \psi
|\psi \rangle =\sum_{j,k}s_{j}^{\ast }s_{k}\langle \alpha _{j}|\alpha
_{k}\rangle =1$ (i.e., $\psi $ is normalized and is a pure state vector of
the field). Given a function $\alpha $, the coherent state $|\alpha \rangle $
of a continuous-mode field is given by the displacement or Weyl operator $%
D(\alpha )$ applied to the vacuum state of the continuous-mode field: 
\begin{equation}
|\alpha \rangle =D(\alpha )|0\rangle .
\end{equation}
The inner product of two coherent states $|\alpha \rangle $, $|\beta \rangle 
$ in the Fock space is given by 
\begin{equation}
\langle \alpha |\beta \rangle =\exp \Big(-\frac{1}{2}\parallel \alpha
\parallel ^{2}-\frac{1}{2}\parallel \beta \parallel ^{2}+\langle \alpha
, \beta \rangle \Big),
\end{equation}
where $\langle g,f \rangle = \int_{-\infty}^{\infty} g(s)^* f(s)ds$ and $\parallel \!\cdot \!\parallel= \langle \cdot
,\cdot \rangle$ are the $L^{2}$ inner product and norm, respectively.
The normalization condition for the superposition state (\ref{eq:super-state})
means that the coefficients must satisfy $\sum_{j,k}s_{j}^{\ast
}s_{k}g_{jk}=1$, where $g_{jk}=\langle \alpha _{j}|\alpha _{k}\rangle $.

More generally, we may consider a field density operator 
\begin{equation}
\rho _{\mathrm{field}}=\sum_{jk}\gamma _{kj}|\alpha _{j}\rangle \langle \alpha _{k}|,
\label{eq:super-state-2}
\end{equation}
that generalizes the superposition state $|\psi \rangle $ to allow for
statistical combinations of coherent states. The normalization for the state $%
\rho _{\mathrm{field}}$ is $\sum_{j,k}\gamma _{jk}g_{jk}=1$.

In what follows the action of the quantum noises $dB$ and $d\Lambda$ on
coherent states will be important: 
\begin{eqnarray}
dB(t) \vert \alpha \rangle &=& \alpha(t) \vert \alpha \rangle dt,  \notag \\
d\Lambda(t) \vert \alpha(t) \rangle &=& dB^\ast(t) \alpha(t) \vert \alpha
\rangle .
\end{eqnarray}


\subsection{Master Equation for Systems Driven by a Field in a Combination or
Superposition of Coherent States}

\label{sec:cat-master} 

Again, before we derive the master equation we introduce some notation that
helps to formulate the master equation. Recall that we defined the
asymmetric expectation $\mathbb{E}_{jk}\left[ {X\otimes F}\right] \equiv
\langle \eta |X|\eta \rangle \langle \phi _{j}|F|\phi _{k}\rangle $. In
section~\ref{sec:photon} we took the field states $|\phi _{j}\rangle ,|\phi
_{k}\rangle $ to be either vacuum or one photon. In this section we use this
same notation but the field states are understood to be continuous-mode coherent
states i.e. $|\alpha _{j}\rangle ,|\alpha _{k}\rangle $. The indices $j,k$
now take the values $1,\ldots ,n$.

The expectation of an arbitrary system observable, with respect to the state 
$|\eta \rangle \langle \eta |\otimes \rho _{\mathrm{field}}$, at time $t$ is 
\begin{equation}
\varpi _{t}(X)=\mathbb{E}_{\eta \rho _{\mathrm{field}}}[X(t)].  \label{mu_super}
\end{equation}
Using the notation (similar to the single photon case) 
\begin{equation}
\varpi _{t}^{jk}(X)=\mathbb{E}_{jk}\left[ {X(t)}\right] =\langle \eta 
\alpha_{j}|X(t)|\eta \alpha_{k}\rangle ,  \label{eq:mujk-super-def}
\end{equation}
with $\rho_{\mathrm{field}}$ as given in \ref{eq:super-state-2}, we may write ~(\ref{mu_super}) as 
\begin{equation}
\varpi _{t}(X)=\sum_{jk}\gamma _{jk}\varpi _{t}^{jk}(X).  \label{notation}
\end{equation}

As in section~\ref{sec:photon-master}, we can derive the Heisenberg master
equation by taking the expectation of the equation of motion for an
arbitrary system operator $dX(t)$, i.e. ~(\ref{eq:qsde-X-1-S}). Doing so
yields the equations 
\begin{equation}
\dot{\varpi}_{t}^{jk}(X)=\varpi _{t}^{jk}(\mathcal{G}_{t}^{jk}X),
\label{eq:dot-mu-jk-c}
\end{equation}
where we define a new superoperator 
\begin{eqnarray}
\mathcal{G}_{t}^{jk}X &\equiv &\mathcal{L}X+S^{\dag }[X,L]\alpha _{j}^{\ast
}(t)+[L^{\dag },X]S \alpha_{k}(t)  \notag \\
&&+(S^{\dag }XS-X)\alpha _{j}^{\ast }(t)\alpha _{k}(t),  \label{admeprop}
\end{eqnarray}
with initial conditions $\varpi _{0}^{jk}(X)=\langle \eta |X|\eta \rangle
g_{jk}$. Note that equations (\ref{eq:dot-mu-jk-c}) are uncoupled.

The corresponding density operator is 
\begin{equation}
\varrho (t)=\sum_{jk}\gamma _{jk}\varrho^{jk}(t)  \label{eq:s-super-ddt-1}
\end{equation}
where 
\begin{eqnarray}
\dot{\varrho}^{jk}=\mathcal{G}_{t}^{jk\star}[\varrho^{j k} ] &\equiv &\mathcal{L}%
^{\star }\varrho +[S\varrho^{jk} ,L^{\dag }]\alpha _{j}(t)+[L,\varrho S^{ \dag}]\alpha _{k}^{\ast
}(t)  \notag \\
&&+(S\varrho^{jk} S^{\dag }-\varrho^{jk} )\alpha _{j}(t)\alpha _{k}^{\ast }(t),
\label{s-admeprop}
\end{eqnarray}
and $\varrho ^{jk}(0)=\left| {\eta }\right\rangle \!\left\langle {\eta }\right| 
{g_{jk}}$. 

The master equations (\ref{eq:dot-mu-jk-c}) and (\ref{s-admeprop}) consist
of a weighted sum of cross-expectations. Clearly these equations reduce to
the vacuum master equation if the only term in the superposition or combination 
is the vacuum.


\subsection{Extended System}

\label{sec:cat-signal} 

In this section we describe a cascade extended system $G_{T}=G\triangleleft
M $ that will be used in section \ref{sec:cat-filter} to determine the
quantum filtering equations for the mixed or superposition of coherent state
field. The ancilla system $M$ will be an $n$-level system, with orthonormal
basis $|j\rangle $, $j=1,\ldots ,n$. The parameters for this system are 
\begin{equation*}
M=(I,L_{M},0),
\end{equation*}
where 
\begin{equation}
L_{M}=\sum_{j}\alpha _{j}(t)|j\rangle \langle j|,
\end{equation}
and we take the initial state of the ancilla to be the density matrix 
\begin{equation}
\rho _{a}=\frac{1}{N_a}\sum_{jk}\gamma _{kj}|j\rangle \langle k|,
\end{equation}
where $N_a=\sum_{l}\gamma _{ll}$ is a normalization factor. The extended
system is 
\begin{equation*}
G_{T}=G\triangleleft M=(S,L+SL_{M},H+\mathrm{Im}\left\{ L^{\dag
}SL_{M}\right\} ).
\end{equation*}

Define $Q_{jk}=|j\rangle \langle k|$. Then a straightforward calculation
shows that 
\begin{equation}
\mathcal{L}_{L_{M}}(Q_{jk})=m_{jk}(t)Q_{jk},
\end{equation}
where 
\begin{equation}
m_{jk}(t)=\alpha _{j}^{\ast }(t)\alpha _{k}(t)-\frac{1}{2}|\alpha
_{j}(t)|^{2}-\frac{1}{2}|\alpha _{k}(t)|^{2}.
\end{equation}

Now consider the extended system $G_{T}$ initialized in the state $\rho
_{a}\otimes |\eta \rangle \langle \eta |\otimes |0\rangle \langle 0|$
(driven by vacuum $|0\rangle $). Then the methods used in Sections \ref
{sec:extended-photon} and \ref{sec:photon-mixed} may be adapted to the
present case to show that 
\begin{equation}
\varpi _{t}^{jk}(X)=\frac{\tilde{\varpi}_{t}(Q_{jk}\otimes X)}{w_{jk}(t)},
\label{eq:claim-mujk-cat}
\end{equation}
where $w_{jk}(t)$ is defined to be the solution of 
\begin{equation}
\dot{w}_{jk}(t)=m_{jk}(t)w_{jk}(t),\ \ w_{jk}(0)=\frac{1}{N_ag_{jk}},
\label{eq:wjk-ode}
\end{equation}
and 
\begin{equation}
\mathbb{E}_{\eta \rho _{\mathrm{field}}}[X(t)]=\mathbb{E}_{\rho _{a}\eta 0}[\tilde{U}%
^{\dag}(t)(R(t)\otimes X)\tilde{U}(t)],  \label{eq:super-represent-1}
\end{equation}
where 
\begin{equation}
R(t)=\sum_{j,k}\frac{\gamma _{jk}}{w_{jk}(t)}Q_{jk},
\end{equation}
These expressions are very similar to the photon case, but with some
important differences. For instance, the ancilla was initialized in the
excited state for the photon case, while here for the mixed coherent case
the initial ancilla state is the density $\rho _{a}$.


\subsection{The Stochastic Master Equation (Filter) for Amplitude Quadrature
Measurements}

\label{sec:cat-filter} 

The quantum filter for the general combination of coherent state case may
now be derived in exactly the same way as was done for the combination of
single photon and vacuum in Section \ref{sec:photon-mixed}. The conditional
expectation we are interested in is 
\begin{equation}
\pi _{t}(X)=\mathbb{E}_{\eta \rho _{\mathrm{field}}}[X(t)\,|\,Y(s),0\leq s\leq t],
\end{equation}
where now $\rho _{\mathrm{field}}$ is given by (\ref{eq:super-state-2}). Equations (\ref
{eq:bayes-R}), (\ref{eq:photon-mixed-filter}), and (\ref
{eq:photon-cdl-density}) again hold, but with modifications to the terms as
described above. The filtering equations are as follows.

The conditional quantities $\pi _{t}^{jk}(X)$ satisfy the coupled system of
equations 
\begin{equation*}
d\pi _{t}^{jk}(X)=\pi _{t}^{jk}(\mathcal{G}^{jk}X)dt+\mathcal{H}%
_{t}^{jk}\left( X\right) dW(t)
\end{equation*}
where the innovations process $W(t)$ is a Wiener process and is given by 
\begin{align*}
dW(t)=& dY(t) \\
& -\sum_{l}\frac{\gamma _{ll}}{N_a}\pi _{t}^{ll}(L+S \alpha%
_{l}(t)+L^{\dag }+S^{\dag }\alpha _{l}^{\ast }(t))dt.
\end{align*}
and the new superoperator $\mathcal{H}_{l}^{jk}\left( \cdot \right) $ is
defined by 
\begin{align*}
&\lefteqn{\mathcal{H}_{l}^{jk}\left( X\right)}\\
&\!\equiv \pi _{t}^{jk}\big (X(L+S%
\alpha_{k}(t))+(L^{\dag }+S^{\dag }\alpha _{j}^{\ast}(t))X\big ) \\
&\quad \!-\pi _{t}^{jk}(X)\sum_{l}\frac{\gamma _{ll}}{N_a}\pi ^{ll}({L+L^{\dag
}+S \alpha_{l}(t)+S^{\dag }\alpha _{l}^{\ast}(t))}.
\end{align*}
As before, we may write $\pi _{t}^{jk}(X)=\mathrm{tr}\left\{ {\varrho }%
^{jk}(t)^{\dag}X\right\} $, where $\varrho ^{jk}(t)$ satisfies the coupled
differential equations (for $j,k=1,2,\ldots ,n$): 
\begin{equation}
d\rho ^{jk}(t)=\mathcal{G}_{t}^{jk\star }[\rho ^{jk}(t)]dt+\mathcal{H}%
_{t}^{jk\star }\left[ {\rho ^{jk}(t)}\right] dW(t)
\label{eq:cat-density-cdl}
\end{equation}
where 
\begin{align}
\mathcal{H}_{t}^{jk\star }\left[ {\rho ^{jk}}\right] \equiv & (L+S
\alpha_{k}(t))\rho ^{jk}+\rho ^{jk}(L^{\dag }+S^{\dag }\alpha _{j}^{\ast
}(t)) \notag \\
 -\rho ^{jk}\sum_{l}&\frac{\gamma _{ll}}{N_a}\mathrm{tr}\left[ {{%
(L+L^{\dag }+S \alpha_{l}(t)+S^{\dag }\alpha _{l}^{\ast}(t))\rho ^{ll}}}%
\right] ,  \notag
\end{align}
with initial conditions $\rho _{0}^{jk}(t)=|\eta \rangle \langle \eta
|g_{jk}$ (recall that $g_{jk}=\langle \alpha _{j}|\alpha _{k}\rangle $). The
conditional density operator is given by (\ref{eq:photon-cdl-density}), with
the $\rho ^{jk}(t)$ given instead by (\ref{eq:cat-density-cdl}).

We remark that the innovations for the cat case now depends on the weights,
in contrast to the mixed photon/vacuum case. 


\subsection{The Stochastic Master Equation (Filter) for Photon Counting
Measurements}

\label{sec:super-counting} 
Analogously, we may also compute the quantum filtering equations for a
system driven by a coherent superposition in the case where the measurement
performed on the output field, $Y(t)$, is photon counting. The filtering
equations in the Heisenberg form are given by (for $j,k=1,2,\ldots,n$): 
\begin{widetext}
\begin{eqnarray*}
d \pi^{jk}_t(X) &=& \pi^{jk}_t( \mathcal{G}^{jk}(X) )dt + \biggl(\frac{\pi_t^{jk}(L^\dag XL + \alpha_k(t)  L^\dag XS +\alpha_j^*(t) S^\dag XL +\alpha_j^*(t)\alpha_k(t)  S^\dag XS)} { \sum_{j=1}^n \frac{\gamma_{jj}}{N_a}  \pi_t^{jj}(L^\dag L+\alpha_j(t)  L^\dag S +\alpha_j^*(t)S^\dag L+ |\alpha_j|^2 I)}   - \pi_t^{jk}(X) \biggr)dN(t),
\end{eqnarray*}
where 
\begin{eqnarray}
dN(t)&=& dY(t)- \sum_{j=1}^n \frac{\gamma_{jj}}{N_a} \pi_t^{jj}(L^\dag L+\alpha_j(t)  L^\dag S +\alpha_j^*(t)S^\dag L + |\alpha_j(t)|^2 I )dt,
\end{eqnarray}
and with initial conditions $\pi_0^{jk}(X) = \langle \eta| X | \eta \rangle g_{jk}$. {The corresponding \sch-picture filter is
\begin{eqnarray*}
d \rho^{jk} (t) &=& \mathcal{G}_{t}^{jk\star }[\rho ^{jk}]dt + \biggl(\mathcal{N}^{-1}[L\rho^{jk}L^\dag + \alpha_k(t) S\rho^{jk}  L^\dag  +\alpha_j^*(t) L \rho^{jk} S^\dag  +\alpha_j^*(t)\alpha_k(t)  S \rho^{jk} S^\dag ) ]   - \rho^{jk} \biggr)dN(t),
\end{eqnarray*}
where 
\begin{align}
\mathcal{N}=\sum_{j=1}^n \frac{\gamma_{jj}}{N_a}  \tr{\rho^{jj}(L^\dag L+\alpha_j(t)  L^\dag S +\alpha_j^*(t)S^\dag L+ |\alpha_j|^2 I)}
\end{align}
and $dN(t)= dY(t)- \mathcal{N}dt$, with initial conditions $\rho^{jk}(0) = \op{\eta}{\eta} g_{jk}$.}
\end{widetext}

\section{Conclusion}

\label{sec_conc}

We have shown that quantum filtering may be extended beyond the Gaussian input situation
to consider a range of non-classical states that are of current interest. Photon wave packet shaping 
is already being applied experimentally and our filtering equations for the single photon input
completes the problem addressed by Gheri et al. in \cite{GheEllPelZol99} by giving the 
quantum trajectories associated to the master equation they derive. We extend this general combinations
of the vacuum an a one photon state through a straightforward weighting procedure. The filter
equations themselves have potential applications to areas such as shaping wave packet for
maximal / minimal absorption by, for instance, a two level atom, or to controlling the system so as
to shape the outgoing field.

We have also derived the quantum filter for cat states. While the concept of an environment
being in a superposition of states may seem unphysical from the perspective of macroscopic 
superselection rules, as we have seen this may effectively be what happens internally once
a standard input is first fed through an appropriate filter system $M$. This leads naturally to 
questions of decoherence \cite{Kupsch97}, and whether preparing input in a cat state is
advantageous in preventing decoherence of cat states for a given system. It is now
experimentally possible to isolate quantum systems sufficiently well to create cat
states in a laboratory \cite{NeeNieHet06,OurTuaLau06,OurHyuTua07}. The cat-state filtering equation
will be of importance for investigating questions as to whether such superpositions may
protected via appropriate environment engineering. 

\emph{Acknowledgements.} The authors wish to thank J.~Hope for helpful
discussions and for pointing out reference \cite{single-photon-production} to us. 
We also wish
to thank A.~Doherty, H.~Wiseman, E.~Huntington and an anonymous referee (of
an earlier version of this manuscript) for helpful discussions and
suggestions; G. Zhang for carefully reading and earlier version of this
manuscript; {and B. Baragiola for discussions about sec. III H}. MJ and HIN gratefully acknowledge the support of the Australian
Research Council. JC acknowledges support from National Science Foundation
Grant No. PHY-0903953 and Office of Naval Research Grant No. N00014-11-1-008.
JG gratefully acknowledges the support of the UK Engineering and Physical Sciences
Research Council through Research Project EP/H016708/1

\label{sec:app-bayes}

\end{document}